\documentclass[a4paper,11pt]{article}

\usepackage[left=2.5cm,right=2.5cm,top=2.5cm,bottom=2.5cm]{geometry}
\usepackage{graphicx,amssymb,amsmath,amsthm,mathrsfs,setspace,subcaption,cite,authblk,float,cancel}

\usepackage[colorlinks=true,citecolor=blue,urlcolor=blue,bookmarks=true]{hyperref}

\usepackage{mathtools}

\DeclarePairedDelimiterX\braket[2]{\langle}{\rangle}{#1 \delimsize\vert #2}

\newcommand{\dif}{\mathrm{d}}
\newcommand{\Eqref}[1]{(\ref{#1})}
\newcommand{\half}{\frac{1}{2}}
\newcommand{\expo}[1]{\mathrm{e}^{#1}}
\newcommand{\brac}[1]{\left(#1 \right)}
\newcommand{\sbrac}[1]{\left[#1\right]}

\numberwithin{equation}{section}
 
\begin{document}

\title{Magnetising de Sitter and Anti-de Sitter spacetimes}

\author{Yun-Ten Chin\footnote{Email: yuntenchin@gmail.com} }

\author{Yen-Kheng Lim\footnote{Email: yenkheng.lim@gmail.com, yenkheng.lim@xmu.edu.my}}

\affil{\normalsize{\textit{School of Mathematics and Physics, Xiamen University Malaysia, 43900 Sepang, Malaysia}}}

\date{\normalsize{\today}}

\renewcommand\Authands{, and }

\maketitle

\begin{abstract}
  We attempt to magnetise the de Sitter (dS) and Anti-de Sitter (AdS) spacetimes in spherical symmetry. First, the weak field case is considered where Maxwell's equation is solved to find magnetic fields in fixed dS/AdS backgrounds. For strong magnetic fields we consider Einstein--Maxwell gravity with a gravitating fluid source. With gravitational backreaction of the magnetic field taken into account, the strong fields deform the dS/AdS spacetime, resulting in a dS/AdS-type analogue of the Melvin magnetic universe. This solution is obtained via a Harrison-like transformation, along with appropriate transformations of the fluid energy and pressures. Some of its physical and geometrical properties of the solution are studied.
\end{abstract}

\section{Introduction} \label{sec_intro}

The Melvin spacetime \cite{Melvin:1963qx} is an exact solution in Einstein--Maxwell gravity describing an axi-symmetric bundle of magnetic field lines held together under its own gravity. This solution, along with the Reissner--N\"{o}rdstrom black hole, are perhaps among the simplest exact solutions that describe the combined effects of gravitational and electromagnetic fields on a spacetime. While it is often called the \emph{Melvin solution} or \emph{Melvin universe}, it should be noted that there are earlier related works due to Bonnor \cite{Bonnor:1954}, and Misra and Radhakrishna \cite{misra1962some}.

Aside from directly solving the Einstein--Maxwell equations, the Melvin spacetime is more elegantly derived by Harrison's solution-generating procedure \cite{Harrison:1968wue} applied to a vacuum Minkowski seed. When applied to the Schwarzschild or Kerr spacetimes, one obtains the corresponding black hole solutions in the Melvin universe \cite{Ernst:1976mzr}. The Harrison transformation method has since been extended to include dilatonic (scalar) fields \cite{Dowker:1993bt,Radu:2003av,Agop:2005np,Lim:2026cky}, higher-dimensional spacetimes \cite{Ortaggio:2004kr,Yazadjiev:2005gs}, and dynamical external fields \cite{BenAchour:2026zui}, and various other directions \cite{Golubtsova:2009esx,Bolokhov:2019jjf,Bolokhov:2019ktk,Bini:2022xzk,Bouzenada:2024ryh,Castro:2024ayd,Tsagas:2020lal}. The scalar analogue to the Melvin solution was also given in Einstein-scalar theory \cite{Cardoso:2024yrb,Herdeiro:2024oxn}.

Beyond magnetising flat spacetimes, a subsequent avenue to explore is the possibility of finding analogous solutions by magnetising the de Sitter (dS) or Anti-de Sitter (AdS) spacetimes. As solutions of Einstein gravity with a cosmological constant $\Lambda$, this seems more of a challenge because the Harrison transformation procedure fails when a cosmological constant is present in the Einstein--Maxwell equations. Nevertheless, various related solutions in Einstein--Maxwell gravity with a non-zero $\Lambda$ has been found \cite{Zofka:2019yfa,Vesely:2019ajp,Vesely:2021jlc,Vesely:2022vws,Ahmed:2025ohc,Bouzenada:2024ryh,Castro:2024ayd}. Of particular relevence to the present work is the one obtained by Astorino \cite{Astorino:2012zm} via a suitable generalisation of the Ernst potentials. The same solution can also be obtained as a limit of the charged C-metric \cite{Havrdova:2006gi,Lim:2018vbq}. When $\Lambda$ is negative, the solution is asymptotically AdS in planar slicing, and the properties are studied in \cite{Kastor:2020wsm}. The properties of the positive $\Lambda$ case was studied in \cite{Toh:2025ykf}.

Despite the progress made by these solutions, a few unaddressed aspects remain. This $\Lambda$-Melvin metric by Astorino contains a part that is conformal to Minkowski, similar to the $\Lambda=0$ solution. Thanks to this property, the solution has a smooth $\Lambda\rightarrow0$ limit recovering the original Melvin spacetime. But because of this, $\Lambda<0$ describes a magnetised version of AdS with planar (i.e., Poincar\'{e}) slicing, hence it does not describe a magnetised counterpart with \emph{spherical} slicing. On the other hand, the metric in the $\Lambda>0$ solution is a warped product of the form $\brac{\mbox{\textsf{Minkowski}}_2}\times S^2$ (in the four-dimensional theory, and $S^2$ is the two-sphere), and does not have the `familiar' features one would expect from a dS spacetime. Namely, (i) it does not have a cosmological horizon, and (ii) the zero field limit is a nakedly singular \cite{Toh:2025ykf}. Furthermore, the planar symmetry makes it challenging to add a black hole of spherical topology into the solution, if one is seeking a suitable generalisation to the Ernst/Schwarzschild--Melvin solution \cite{Ernst:1976mzr}. 

In this paper, we attempt to close this gap by seeking a magnetised version of (A)dS spacetime in spherical symmetry. That is, prior to adding the magnetic field, the spacetime has the metric
\begin{align}
 \dif s^2&=-f(r)\dif t^2+f(r)^{-1}\dif r^2+r^2\dif\Omega^2,\quad f(r)=1-\frac{\Lambda}{3}r^2, \label{eq_PureBackground}
\end{align}
where $\dif\Omega^2=\dif\theta^2+\sin^2\theta\,\dif\phi^2$ is the metric of a unit two-sphere. The cases $\Lambda>0$ and $\Lambda<0$ correspond to the dS and AdS spacetimes, respectively. We will start by first adding a \emph{test} magnetic field. That is, we work in the weak field regime by solving Maxwell's equation with \Eqref{eq_PureBackground} as a fixed background. This assumes the field is sufficiently weak that there is no backreaction to the spacetime geometry. The Maxwell equation is first solved in vacuum (i.e., zero charge/current sources) and find the first few multipole solutions, giving the magnetic analogue to the results of \cite{Herdeiro:2015vaa}. We then use a method inspired by Wald \cite{Wald:1974np} where Killing vectors are used as solutions to the Maxwell equation in Ricci-flat spacetimes. However, in our case the spacetime is not Ricci-flat and a current source proportional to $\Lambda$ is required in the Maxwell equation.

The requirement of a current source then gives a clue that, to fully magnetise the spacetime including gravitational backreaction, we will require the presence of matter. This also circumvents the issue of $\Lambda$ spoiling the symmetry of the Harrison transformation, as now $\Lambda$ need not be regarded as a cosmological \emph{constant}, but rather a fluid with the `dark energy' equation of state. Solution-generating transformations in the presence of matter fluids was previously done in \cite{Yazadjiev:2004bg,Yazadjiev:2011sm} to generate solutions with scalar fields, and Harrison-type transformations by Stelea et al. \cite{Stelea:2018cgm,Lungu:2024iob,Lungu:2025pgk,Al-Badawi:2025abo} to magnetise various black hole solutions. Here, we apply a Harrison-type transformation to produce magnetised versions of the dS and AdS spacetime. In this model, there is no cosmological constant per se. But rather we have a gravitating fluid source that reduces to an effective cosmological constant in the zero field limit.

As such, the main framework used in this paper is Einstein--Maxwell--fluid gravity. In this paper, we use the convention $(-,+,+,+)$ for Lorentzian signature. In lightspeed units where $c=1$ and keeping the other physical constants explicit for the moment, the action is
\begin{align}
 I=\int\dif^4x\sqrt{-g}\brac{\frac{1}{16\pi G }R-\frac{1}{4\mu_0}F^2+\mathcal{L}_{\mathrm{m}}}, \label{action}
\end{align}
where $g=\det g_{\mu\nu}$ and $F^2=F_{\mu\nu}F^{\mu\nu}$ is the self-contraction of the Faraday tensor $F=\dif A$ with components $F_{\mu\nu}=\partial_\mu A_\nu-\partial_\nu A_\mu$, which is the exterior derivative of a gauge potential $A=A_\mu\dif x^\mu$. As we will see below, if $\mathcal{L}_{\mathrm{m}}\propto\Lambda$ is simply the cosmological constant, the Harrison transformation procedure fails. Hence we will take $\mathcal{L}_{\mathrm{m}}$ to be a matter Lagrangian for a current-carrying anisotropic fluid. Variation of the action with respect to $g^{\mu\nu}$ and $A_\mu$ leads to
\begin{subequations}
\begin{align}
 R_{\mu\nu}-\half R g_{\mu\nu}&=8\pi G\brac{\mathcal{T}_{\mu\nu}+T_{\mu\nu}},\label{eq_EE_units}\\
 \nabla_\lambda F^{\mu\lambda}&=\mu_0 J^\mu,\label{eq_ME_units}
\end{align}
\end{subequations}
where the energy-momentum tensors and four-currents are
\begin{align}
 \mathcal{T}_{\mu\nu}&=\frac{1}{\mu_0}\brac{F_{\mu\lambda}{F_{\nu}}^\lambda-\frac{1}{4}F^2g_{\mu\nu}},\quad T_{\mu\nu}=\mathcal{L}_{\mathrm{m}}g_{\mu\nu}-2\frac{\delta\mathcal{L}_{\mathrm{m}}}{\delta g^{\mu\nu}},\quad J^\mu=\frac{\delta\mathcal{L}_{\mathrm{m}}}{\delta A_\mu}. \label{eq_SourceCurrents}
\end{align}

Working in this setup, the rest of this paper is organised as follows. In Sec.~\ref{sec_WeakFields}, we add test magnetic fields to the dS and AdS spacetimes. In Sec.~\ref{sec_eom}, we consider a modified Harrison transformation procedure to derive a solution to the full Einstein--Maxwell--fluid equations which magnetises the dS and AdS spacetimes. Some physical and geometrical properties of the resulting solutions are studied in Sec.~\ref{sec_properties}. This is followed by Sec.~\ref{sec_geod} which is dedicated to studying its equatorial geodesics. Conclusions and closing remarks are given in Sec.~\ref{sec_conclusion}.

\section{Test fields in the (A)dS background} \label{sec_WeakFields}

We first consider adding test (or probe) fields to a fixed (A)dS background. That is, we assume the fields are weak such that its energy does not backreact to the geometry. From Eq.~\Eqref{eq_EE_units}, this is when $8\pi G\mathcal{T}_{\mu\nu}\ll 1$. Then only the Maxwell equation \Eqref{eq_ME_units} determines the electromagnetic fields on a fixed (A)dS background with metric as given in Eq.~\Eqref{eq_PureBackground}. The Ricci tensor of the metric \Eqref{eq_PureBackground} satisfies $R_{\mu\nu}=\Lambda g_{\mu\nu}$.\footnote{In other words, it solves the Einstein equation \Eqref{eq_EE_units} with  $T_{\mu\nu}=-\frac{\Lambda}{8\pi G} g_{\mu\nu}$ and $8\pi G\mathcal{T}_{\mu\nu}$ being negligible.} The spacetime is de Sitter for $\Lambda>0$ and Anti-de Sitter (in spherical slicing) for $\Lambda<0$. In the following we obtain magnetic fields for this background by (i) solving the vacuum Maxwell equations (Sec.~\ref{sec_Maxwell_vac}--\ref{sec_Maxwell_vac_AdS}), and (ii) using a Killing vector to help find a solution (Sec.~\ref{sec_Maxwell_Killing}).

\subsection{Vacuum Maxwell equations} \label{sec_Maxwell_vac}

In the absence of current sources, the Maxwell equation reads
\begin{align}
 \nabla_\mu F^{\mu\nu}=\frac{1}{\sqrt{-g}}\partial_\mu\brac{\sqrt{-g}F^{\mu\nu}}=0, \label{eq_TestMaxwellEqn}
\end{align}
Previously, the authors of Refs.~\cite{Herdeiro:2015vaa,Costa:2015gol} have solved the vacuum Maxwell equation in AdS case ($\Lambda<0$) to obtain \emph{electric} multipole solutions, and analysed its resulting properties. Here, we seek \emph{magnetic} multipoles and consider both positive and negative $\Lambda$.

To start, we take the ansatz
\begin{align}
 A=\Psi(r,\theta)\dif\phi, \label{eq_WeakA_ansatz}
\end{align}
where $\Psi(r,\theta)$ is assumed to depend only on $r$ and $\theta$. Inserting this into Eq.~\Eqref{eq_TestMaxwellEqn}, we find
\begin{align}
 r^2\partial_r\brac{f\partial_r\Psi}+\sin\theta\partial_\theta\brac{\frac{\partial_\theta\Psi}{\sin\theta}}=0.\label{eq_PDE}
\end{align}
We solve this by applying separation of variables. To this end, take $\Psi(r,\theta)=-R(r)\sin\theta\frac{\dif P(\theta)}{\dif\theta}$, where $R$ and $P$ are single-variable functions of $r$ and $\theta$, respectively. We then find the existence of a separation constant $\lambda$ which allows us to split \Eqref{eq_PDE} into two ordinary differential equations. Upon letting $x=\cos\theta$, we get
\begin{subequations}
\begin{align}
 (1-x^2)\frac{\dif^2P}{\dif x^2}-2x\frac{\dif P}{\dif x}+n(n+1) P&=0,\label{eq_LegendreEqn}\\
 r^2\frac{\dif}{\dif r}\brac{f\frac{\dif R}{\dif r}}-n(n+1) R&=0. \label{eq_REqn}
\end{align}
\end{subequations}
We recognise Eq.~\Eqref{eq_LegendreEqn} as the Legendre equation where $\lambda=n(n+1)$, for integer $n$. The solutions are then the Legendre polynomials,
\begin{align}
 P=P_n(x)=P_n(\cos\theta).
\end{align}
The general solution for Eq.~\Eqref{eq_REqn} may be expressed in terms of Heun functions, but they take simpler forms for specific choices of $n$. Each $n$ determines the radial and angular solutions with $R_n(r)$ and $P_n(\cos\theta)$, respectively.

Once the radial and angular solutions are found, the gauge potential is assembled as
\begin{align}
 A=-R_n(r)\sin\theta\frac{\dif P_n(\cos\theta)}{\dif\theta}\dif\phi,
\end{align}
from which one obtains the Faraday tensor by $F_{\mu\nu}=\partial_\mu A_\nu-\partial_\nu A_\mu$. It might be interesting to visualise the magnetic field configuration as seen from the perspective of a static time-like observer. The four-velocity for such an observer is $u^\mu=\brac{f^{-1/2},0,0,0}$, so that $u^\mu u_\mu=-1$ as required for a time-like observer. The magnetic field components are obtained by
\begin{align}
 B_\mu=-\half\epsilon_{\mu\nu\rho\sigma}u^\nu F^{\sigma\rho},\label{eq_Bcomponents}
\end{align}
where $\epsilon_{\mu\nu\rho\sigma}$ is the Levi--Civita tensor oriented such that $\epsilon_{tr\theta\phi}=+\sqrt{-g}$. Raising the indices to $B^\mu$ and taking the spatial components, the magnetic field is then taken to be $\vec{B}=(B^r,B^\theta,B^\phi)$.

The energy density measured by a static observer with four-velocity $u^\mu$ is obtained from
\begin{align}
 \varrho=u^\mu u^\mu \mathcal{T}_{\mu\nu}, \label{eq_EMrho_def}
\end{align}
where $\mathcal{T}_{\mu\nu}$ is the stress-energy tensor due to the electromagnetic fields as given in Eq.~\Eqref{eq_SourceCurrents}. For gauge potentials of the form \Eqref{eq_WeakA_ansatz}, the energy density as seen by a static observer $u^\mu$ is
\begin{align}
 \varrho=u^\mu u^\nu\mathcal{T}_{\mu\nu}=\frac{1}{4\mu_0}F^2=\frac{1}{2\mu_0 r^2\sin^2\theta}\sbrac{\brac{\partial_r\Psi}^2+\frac{1}{r^2}\brac{\partial_\theta\Psi}^2}. \label{eq_WeakVac_rho}
\end{align}

\subsection{Vacuum weak fields in de Sitter spacetime} \label{sec_Maxwell_vac_dS}

In this case, $\Lambda$ is positive and we write the function $f$ in \Eqref{eq_PureBackground} as
\begin{align}
 f(r)=1-\frac{r^2}{\ell^2},
\end{align}
where $\ell=\sqrt{3/\Lambda}$ is the dS length scale. There is a cosmological horizon at $r=\ell$, and in the following we consider fields in the static patch $0<r<\ell$.

\begin{figure}
 \centering
 \begin{subfigure}[b]{0.49\textwidth}
    \centering
    \includegraphics[width=0.8\textwidth]{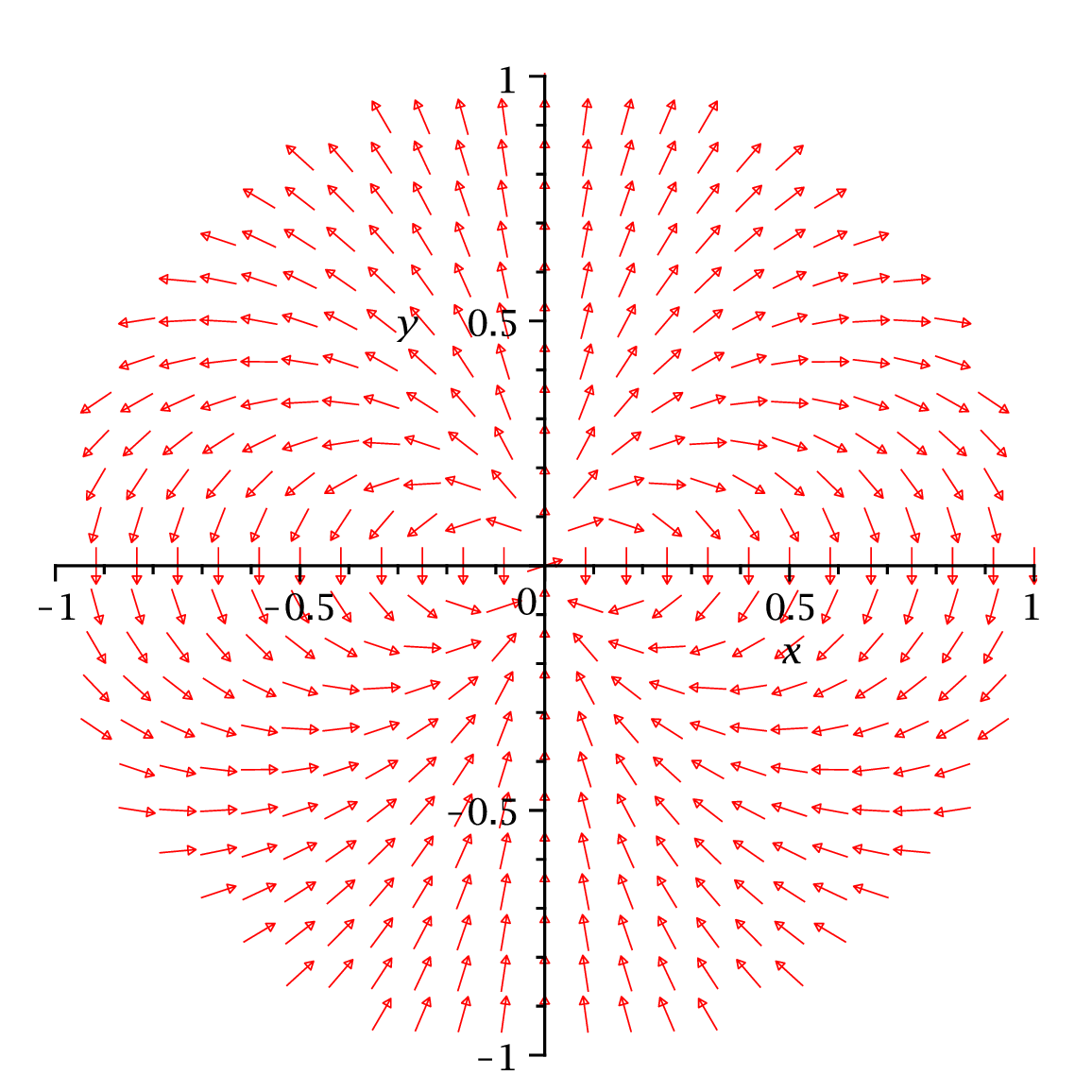}
    \caption{$n=1$.}
    \label{fig_vacB_dS1}
  \end{subfigure}
  \begin{subfigure}[b]{0.49\textwidth}
    \centering
    \includegraphics[width=0.8\textwidth]{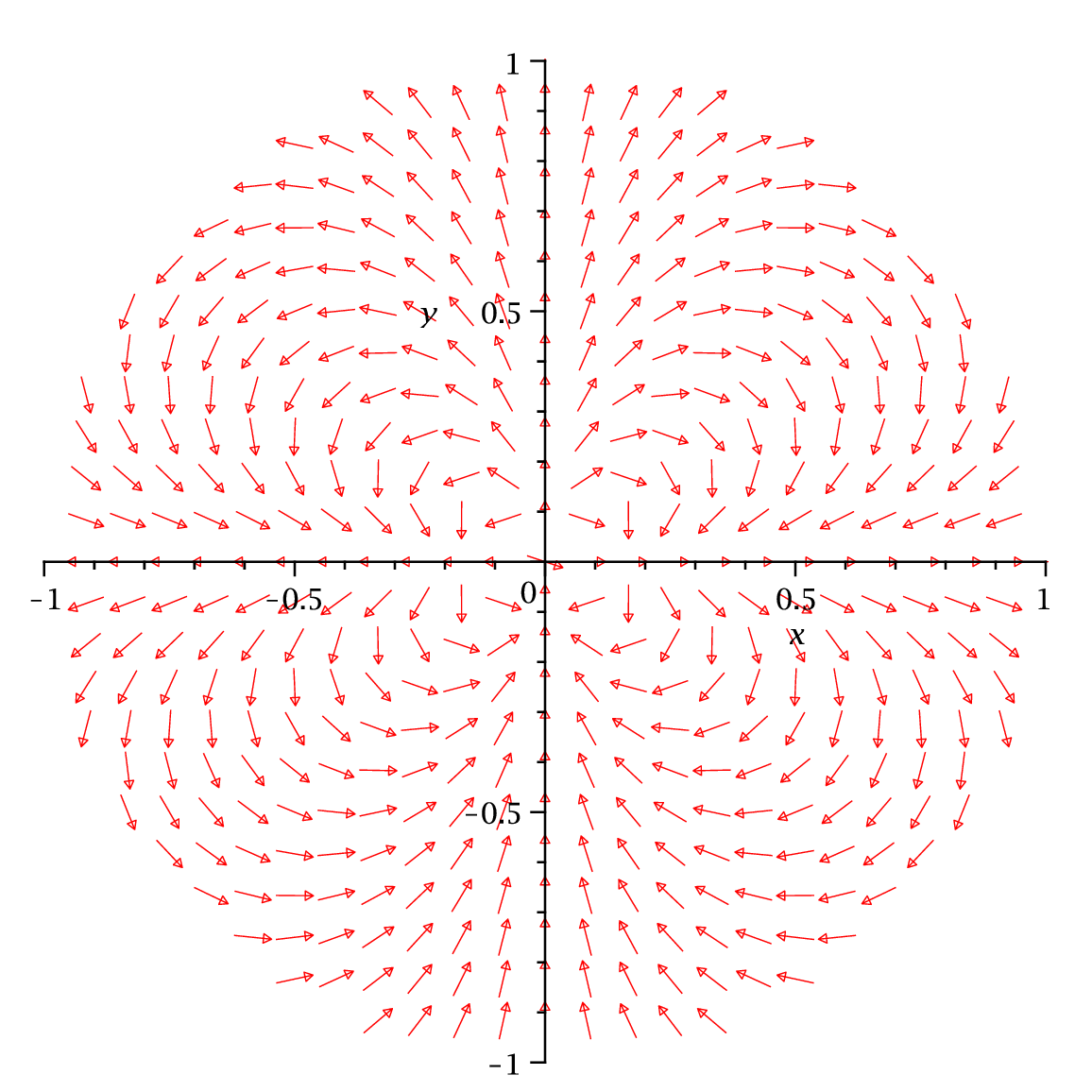}
    \caption{$n=2$.}
    \label{fig_vacB_dS2}
  \end{subfigure}
  \begin{subfigure}[b]{0.49\textwidth}
    \centering
    \includegraphics[width=0.8\textwidth]{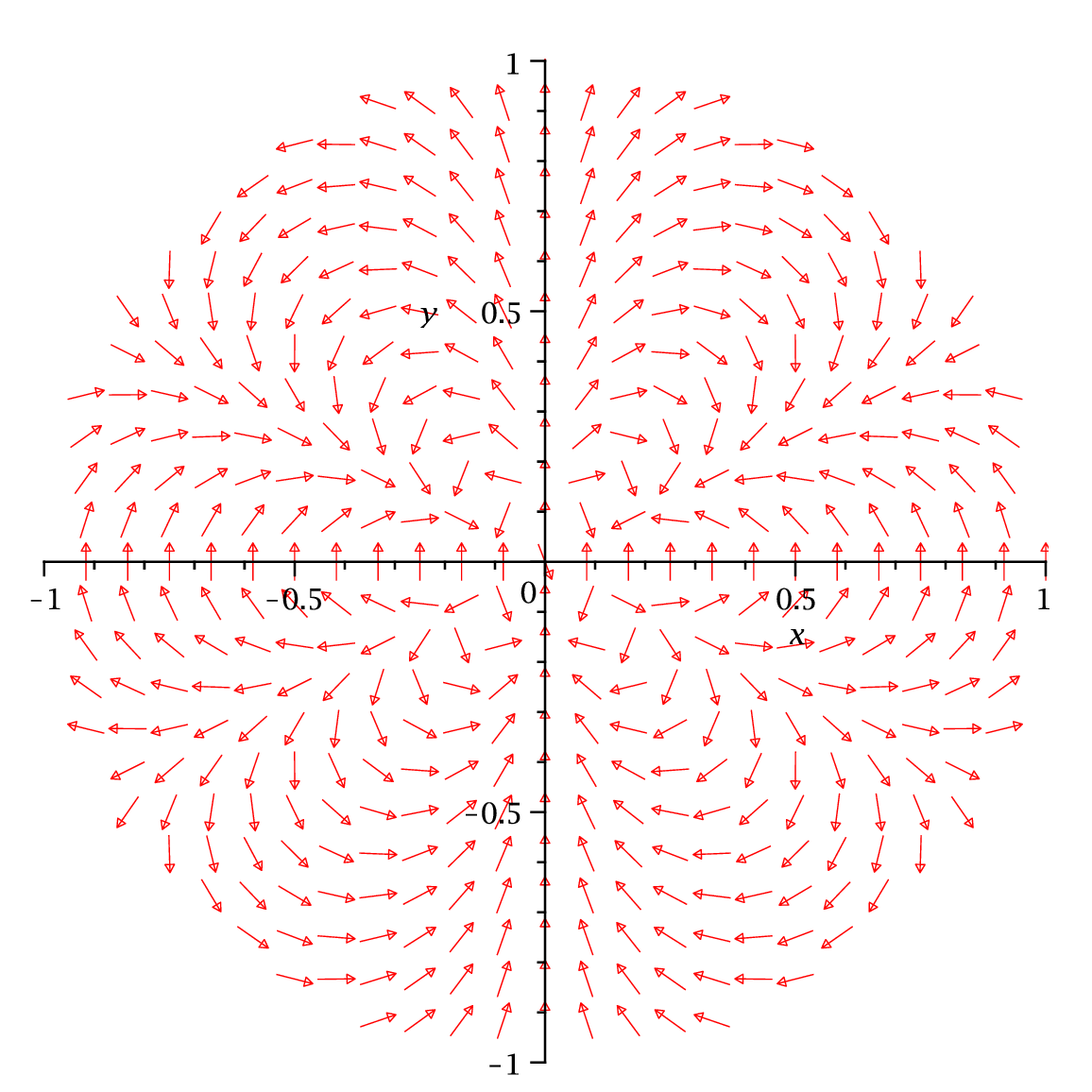}
    \caption{$n=3$.}
    \label{fig_vacB_dS3}
  \end{subfigure}
  \begin{subfigure}[b]{0.49\textwidth}
    \centering
    \includegraphics[width=0.8\textwidth]{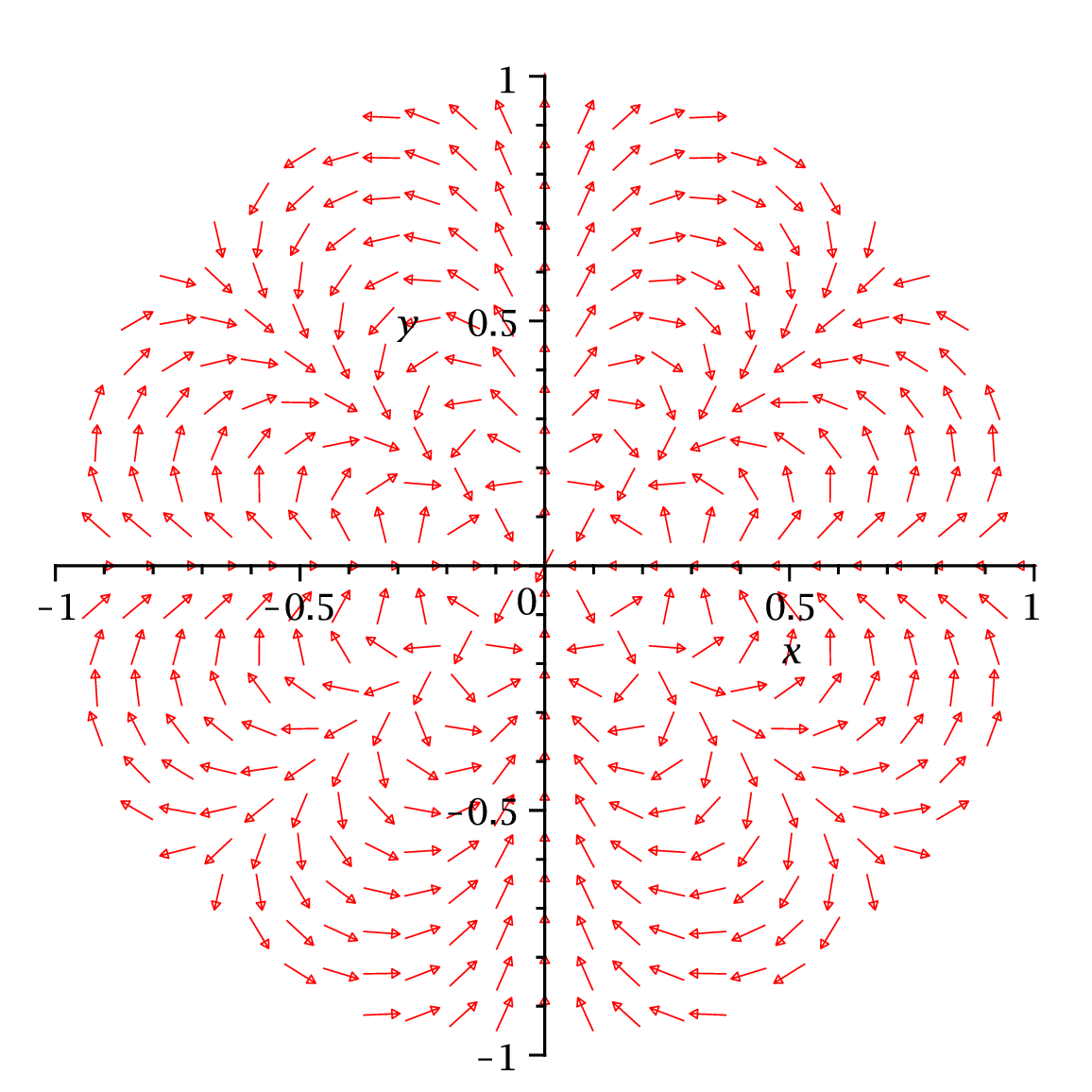}
    \caption{$n=4$.}
    \label{fig_vacB_dS4}
  \end{subfigure}
  \caption{Plots of the spatial magnetic field configuration of the solutions \Eqref{eq_vacB_dSsoln}. The axes are given in units of $\ell$. The fields are only plotted within the static patch $r<\ell$.}
  \label{fig_vacB_dS}
\end{figure}

The solution for $n=1$ is
\begin{align*}
 \Psi=\sbrac{\frac{C_1}{r}+C_2\brac{\frac{\ell}{r}\ln\brac{\frac{\ell+r}{\ell-r}}-2}}\sin^2\theta,
\end{align*}
where $C_1$ and $C_2$ are the integration constants. We note that the part $\frac{\ell}{r}\ln\frac{\ell+r}{\ell-r}$ diverges at the dS horizon, so we set $C_2=0$, and $C_1=b$ so that the solution for the gauge potential is
\begin{align}
 \Psi=\frac{b}{r}\sin^2\theta.
\end{align}
Similarly for the next few $n$, we set one of the integration constants to zero to remove solutions singular at the horizon, and we denote the other constant as $b$. The solutions for the first few $n$ are
\begin{subequations} \label{eq_vacB_dSsoln}
\begin{align}
 n=1:\quad \Psi&=\frac{b}{r}\sin^2\theta,\\
 n=2:\quad \Psi&=3b\brac{\frac{3\ell^2}{r^2}-1}\sin^2\theta\cos\theta,\\
 n=3:\quad \Psi&=b\brac{\frac{3r^2-5\ell^2}{r^3}}\brac{\frac{3}{2}\sin\theta-\frac{15}{2}\cos^2\theta\sin\theta},\\
 n=4:\quad \Psi&=b\brac{\frac{35\ell^4-30\ell^2r^2+3r^4}{r^4}}\sin\theta\brac{\frac{35}{2}\cos^3\theta\sin\theta+\frac{15}{2}\sin\theta\cos\theta}.
\end{align}
\end{subequations}
In Fig.~\ref{fig_vacB_dS}, we plot the resulting fields from the solutions in \Eqref{eq_vacB_dSsoln} using Eq.~\Eqref{eq_Bcomponents}, where the multipole structures for the different $n$ can be seen.
The energy densities of the first four solutions are obtained using Eq.~\Eqref{eq_WeakVac_rho}, giving
\begin{subequations}
 \begin{align}
  n=1:\quad \varrho&=\frac{b^2\sbrac{(3\ell^2+r^2)\cos^2\theta+\ell^2-r^2}}{4\mu_0\ell^2r^6},\\
  n=2:\quad \varrho&=\frac{9b^2}{2\mu_0 r^8}\sbrac{\brac{9r^4+45\ell^4-18\ell^2r^2}\cos^4\theta-\brac{6r^4+18\ell^4}\cos^2\theta+\brac{3\ell^2-r^2}^2},\\
  n=3:\quad \varrho&=\frac{9b^2}{8\mu_0\ell^2r^{10}}\Big[\brac{4375\ell^6+225 r^6+1125\ell^2r^4-4125\ell^4r^2}\cos^6\theta\nonumber\\
   &\quad+(3375r^2\ell^4-4125\ell^6-855\ell^2r^4-315r^6)\cos^4\theta\nonumber\\
   &\quad+(99r^6+1125\ell^6+207\ell^2r^4-855r^2\ell^4)\cos^2\theta+ 9(\ell^2-r^2)(r^2-5\ell^2)^2\Big],\\
   n=4:\quad \varrho&=\frac{25b^2}{8\mu_0r^{12}}\Big[(360150\ell^4r^4+540225\ell^8+11025r^8-788900\ell^6r^2-44100\ell^2r^6)\cos^8\theta\nonumber\\
    &\quad +(-474600\ell^4r^4-788900\ell^8+1097600\ell^6r^2+50400\ell^2r^6-18900r^8)\cos^6\theta\nonumber\\
    &\quad+(360150\ell^8-474600\ell^6r^2+9990r^8+191700\ell^4r^4-16200\ell^2r^6)\cos^4\theta\nonumber\\
    &\quad+(-1620r^8-44100\ell^8+50400\ell^6r^2-16200\ell^4r^4)\cos^2\theta+9990\ell^4r^4-1620\ell^2r^6\nonumber\\
    &\quad +11025\ell^8-18900\ell^6r^2+81r^8\Big].
 \end{align}
\end{subequations}
We see that the solutions are singular at $r=0$. In fact, as $r$ approaches the origin, the energy density $\varrho$ grows arbitrarily large. Eventually the energy density will be large enough that the fields start distorting the spacetime gravitationally, and the weak field description breaks down if $r$ is too close to zero.

\subsection{Vacuum weak fields in Anti-de Sitter spacetime} \label{sec_Maxwell_vac_AdS}

In this case, as $\Lambda$ is negative it is convenient to write $f(r)$ as
\begin{align}
 f(r)=1+\frac{r^2}{L^2},
\end{align}
where $L=\sqrt{-3/\Lambda}$ is the AdS length scale. As before, for each $n$, there are two linearly independent solutions for $R(r)$ to Eq.~\Eqref{eq_REqn}. One of them is decaying in $r$, the other growing. But unlike the dS case, the growing solution is regular, and tends to a finite value as $r\rightarrow\infty$. This behaviour is also seen in the electric analogue in Ref.~\cite{Herdeiro:2015vaa}. For example, the solution to Eq.~\Eqref{eq_REqn} for $n=1$ and $f=1+r^2/L^2$ is
\begin{align}
 R(r)&=\frac{C_1}{r}+C_2\brac{\frac{L}{r}\tan^{-1}\brac{\frac{r}{L}}-1},
\end{align}
where $C_1$ and $C_2$ are the integration constants. The term with coefficient $C_1$ is clearly singular at $r=0$ and asymptotes to zero as $r\rightarrow\infty$. However the term with coefficient $C_2$ is regular,
\begin{align*}
 \lim_{r\rightarrow0}\brac{\frac{L}{r}\tan^{-1}\brac{\frac{r}{L}}-1}&=0,\\
 \lim_{r\rightarrow\infty}\brac{\frac{L}{r}\tan^{-1}\brac{\frac{r}{L}}-1}&=-1.
\end{align*}
Similar behaviours can be see for the higher $n$ solutions. Let us refer to the two cases as \emph{singular} and \emph{regular} solutions respectively. The singular and regular solutions for the first few $n$ are considered briefly in the following.

\subsection*{Singular solutions}
We start with the singular solutions. For the first few multipoles, they are
\begin{align}
 n=1:\quad\Psi&=\frac{b}{r},\nonumber\\
 n=2:\quad\Psi&=\frac{3b}{r^2}(3L^2+r^2)\sin^2\theta\cos\theta,\nonumber\\
 n=3:\quad\Psi&=\frac{b}{2r^3}\brac{5L^2+3r^2}\sin^2\theta\brac{15\cos^2\theta+3},\nonumber\\
 n=4:\quad\Psi&=\frac{b}{2r^4}(35L^4+30r^2L^2+3r^4)\sin^2\theta\brac{35\cos^3\theta-15\cos\theta}. \label{eq_vacB_AdSdecaysoln}
\end{align}
\begin{figure}
 \centering
 \begin{subfigure}[b]{0.49\textwidth}
    \centering
    \includegraphics[width=0.8\textwidth]{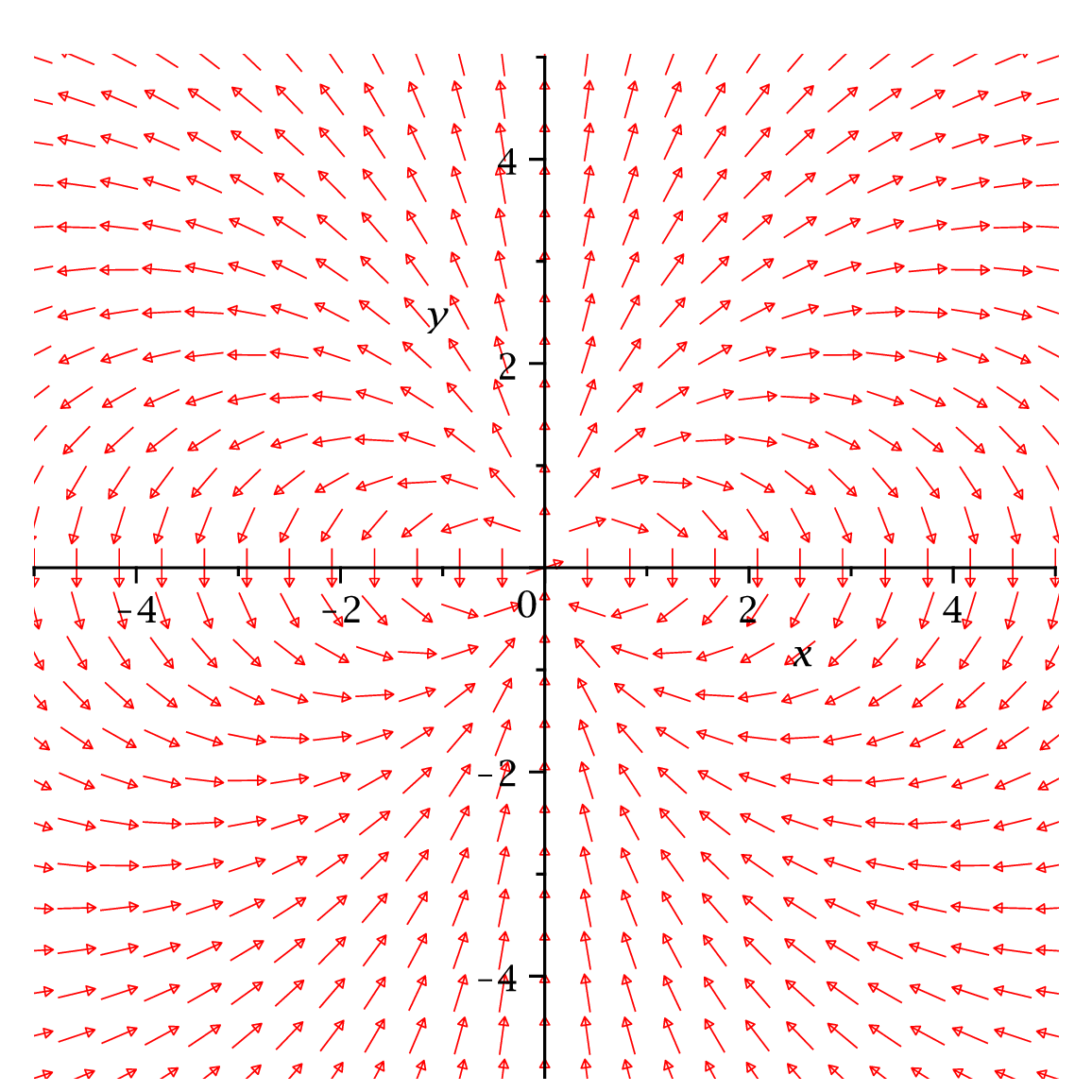}
    \caption{$n=1$.}
    \label{fig_vacB_AdSdecay1}
  \end{subfigure}
  \begin{subfigure}[b]{0.49\textwidth}
    \centering
    \includegraphics[width=0.8\textwidth]{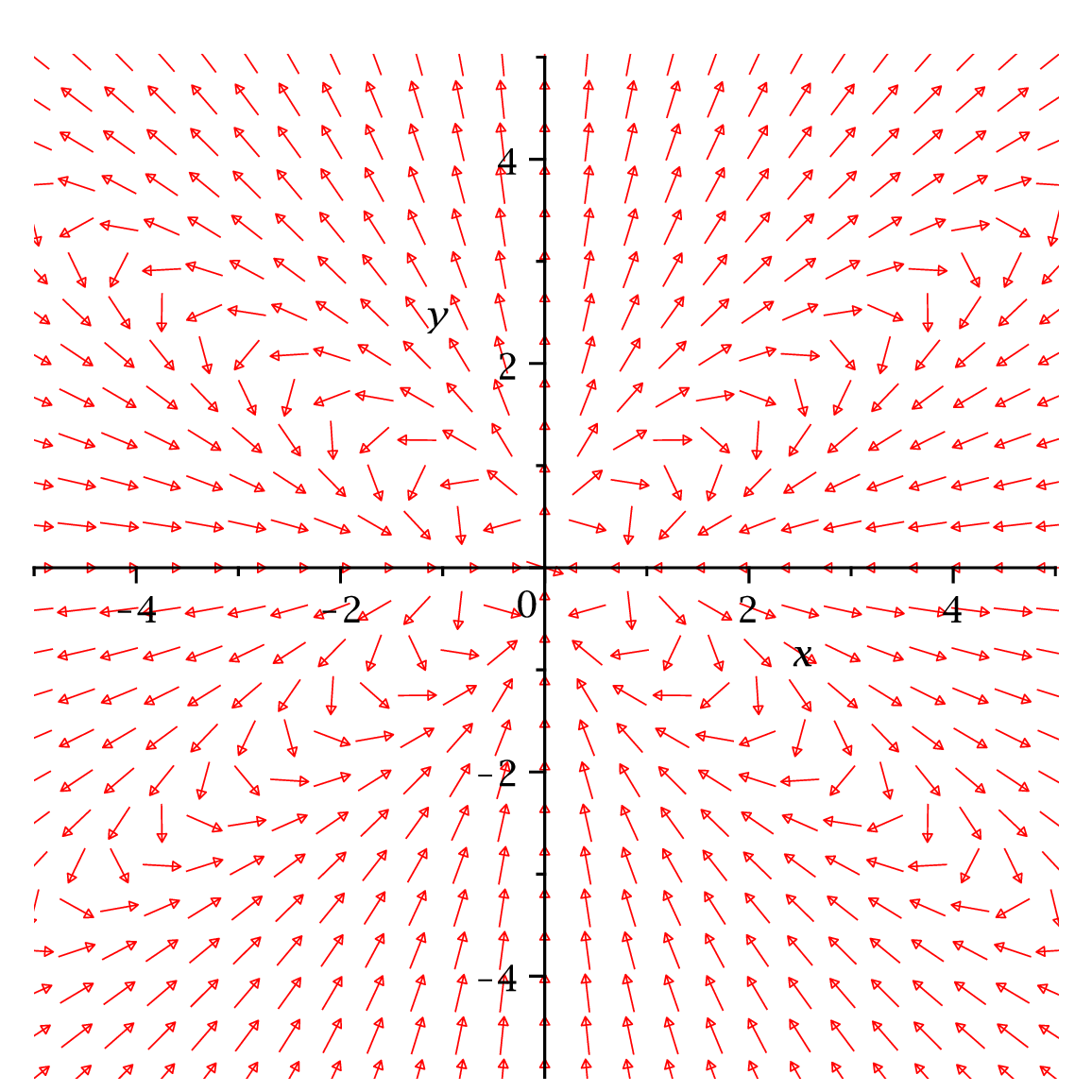}
    \caption{$n=2$.}
    \label{fig_vacB_AdSdecay2}
  \end{subfigure}
  \begin{subfigure}[b]{0.49\textwidth}
    \centering
    \includegraphics[width=0.8\textwidth]{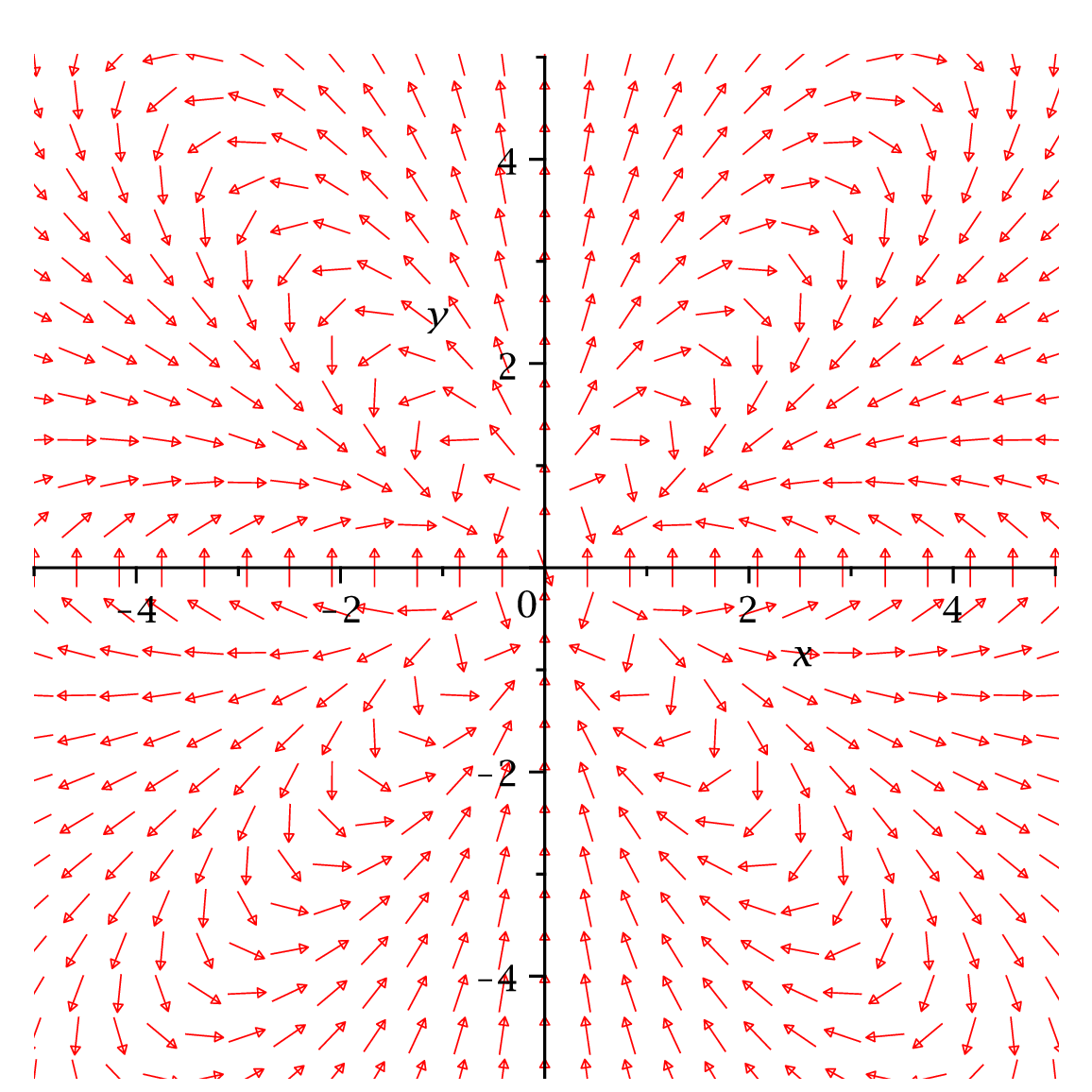}
    \caption{$n=3$.}
    \label{fig_vacB_AdSdecay3}
  \end{subfigure}
  \begin{subfigure}[b]{0.49\textwidth}
    \centering
    \includegraphics[width=0.8\textwidth]{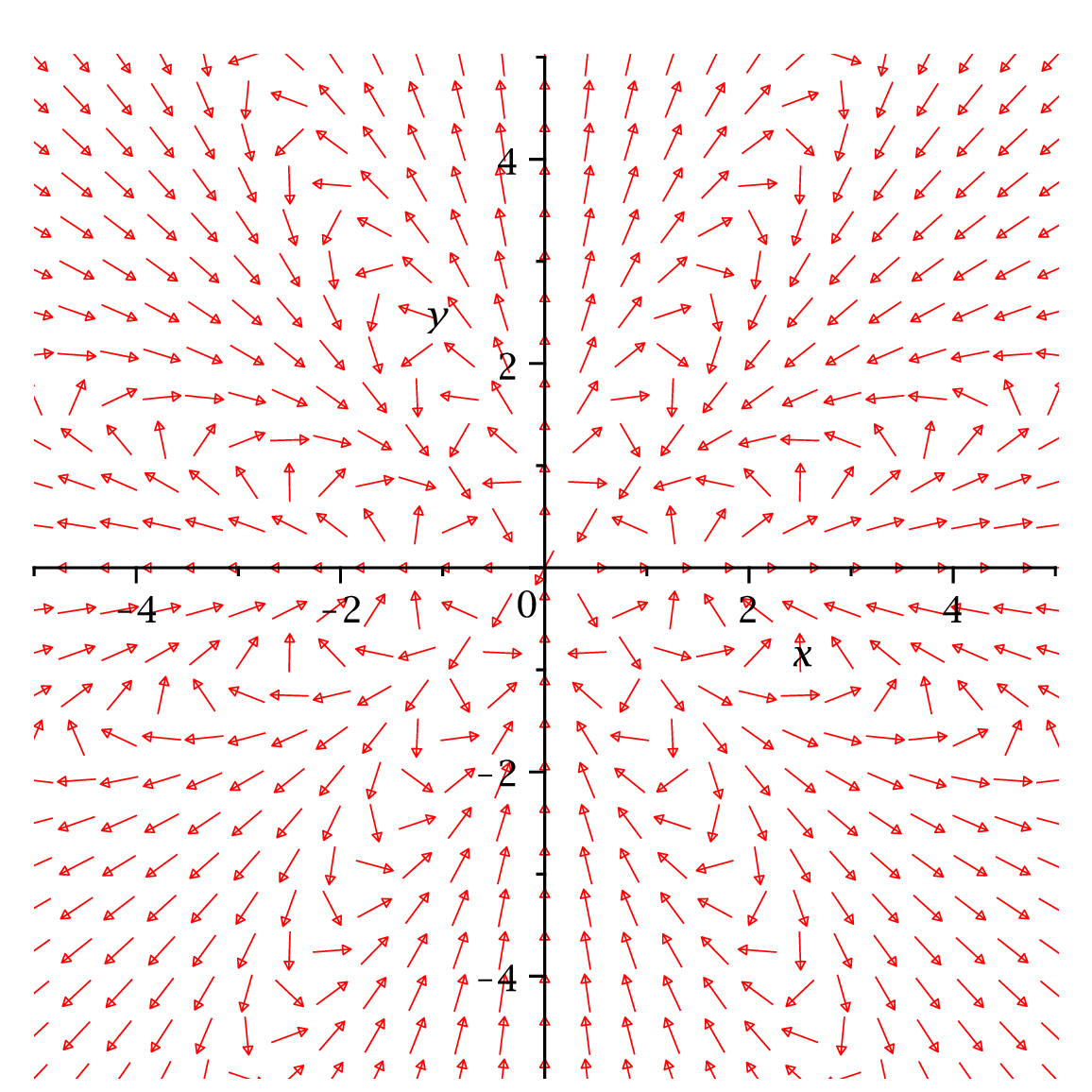}
    \caption{$n=4$.}
    \label{fig_vacB_AdSdecay4}
  \end{subfigure}
  \caption{Plots of the spatial magnetic field configuration of the singular solutions \Eqref{eq_vacB_AdSdecaysoln}. The axes are given in units of $L$.}
  \label{fig_vacB_AdSdecay}
\end{figure}
The energy densities are calculated using Eq.~\Eqref{eq_WeakVac_rho}. For the first four $n$, they are
\begin{align}
 n=1:\quad\varrho&=\frac{b^2}{2\mu_0L^2r^6}\sbrac{(3L^2-r^2)\cos^2\theta+r^2+L^2},\nonumber\\
 n=2:\quad\varrho&=\frac{9b^2}{2\mu_0r^8}\sbrac{(45L^4+18L^2r^2+9r^4)\cos^4\theta-6(3L^4-r^4)\cos^2\theta+(3L^2+r^2)^2},\nonumber\\
 n=3:\quad\varrho&=\frac{9b^2}{4\mu_0L^2r^{10}}\big[(4375L^6-225r^6+4125r^2L^4+1125L^2r^4)\cos^6\theta\nonumber\\
  &\quad+(-3375r^2L^4-855L^2r^4-4125L^6+315r^6)\cos^4\theta\nonumber\\
  &\quad+(207L^2r^4+855r^2L^4-99r^6+1125L^6)\cos^2\theta+9(L^2+r^2)(r^2+5L^2)^2,\nonumber\\
  n=4:\quad\varrho&=\frac{25b^2}{8\mu_0r^{12}}\big[(540225L^8+788900L^6r^2+44100L^2r^6+11025r^8+360150L^4r^4)\cos^8\theta\nonumber\\
  &\quad+(-18900r^8-788900L^8-1097600L^6r^2-474600L^4r^4-50400L^2r^6)\cos^6\theta\nonumber\\
  &\quad+(9990r^8+360150L^8+16200L^2r^6+191700L^4r^4+474600L^6r^2)\cos^4\theta\nonumber\\
  &\quad+(-1620r^8-50400L^6r^2-44100L^8-16200L^4r^4)\cos^2\theta\nonumber\\
  &\quad+9(35L^4+30r^2L^2+3r^4)^2.
\end{align}
Similar to the dS case, they grow arbitrarily large as $r$ approaches zero. Therefore at some point near the origin, the weak field description breaks down. The magnetic field configurations seen by a static observer, using Eq.~\Eqref{eq_Bcomponents} are shown for the first four $n$ in Fig.~\ref{fig_vacB_AdSdecay}.

\subsection*{Regular solutions}
Turning now to the regular solutions, for the first few $n$, they are
\begin{align}
 n=1:\quad\Psi&=b\sbrac{\frac{L}{r}\tan^{-1}\brac{\frac{r}{L}}-1}\sin^2\theta,\nonumber\\
 n=2:\quad\Psi&=3b\sbrac{\brac{1+\frac{3 L^2}{r^2}}\tan^{-1}\brac{\frac{r}{L}}-\frac{3L}{r}}\sin^2\theta\cos\theta, \nonumber\\
 n=3:\quad\Psi&=\frac{b}{8}\sbrac{\brac{\frac{5L^3}{r^3}+\frac{3L}{r}}\tan^{-1}\brac{\frac{r}{L}}-\frac{5L^2}{r^2}-\frac{4}{3}}\sin^2\theta\brac{3-15\cos^2\theta}, \nonumber\\
 n=4:\quad\Psi&=\frac{b}{2r^4}\sbrac{\brac{L^4+\frac{6}{7}L^2r^2+\frac{3}{35}r^4}\tan^{-1}\brac{\frac{r}{L}}}\sin^2\theta\brac{35\cos^3\theta-15\cos\theta}. \label{eq_vacB_AdSgrowingsoln}
\end{align}

\begin{figure}[htb!]
 \centering
 \begin{subfigure}[b]{0.49\textwidth}
    \centering
    \includegraphics[width=0.8\textwidth]{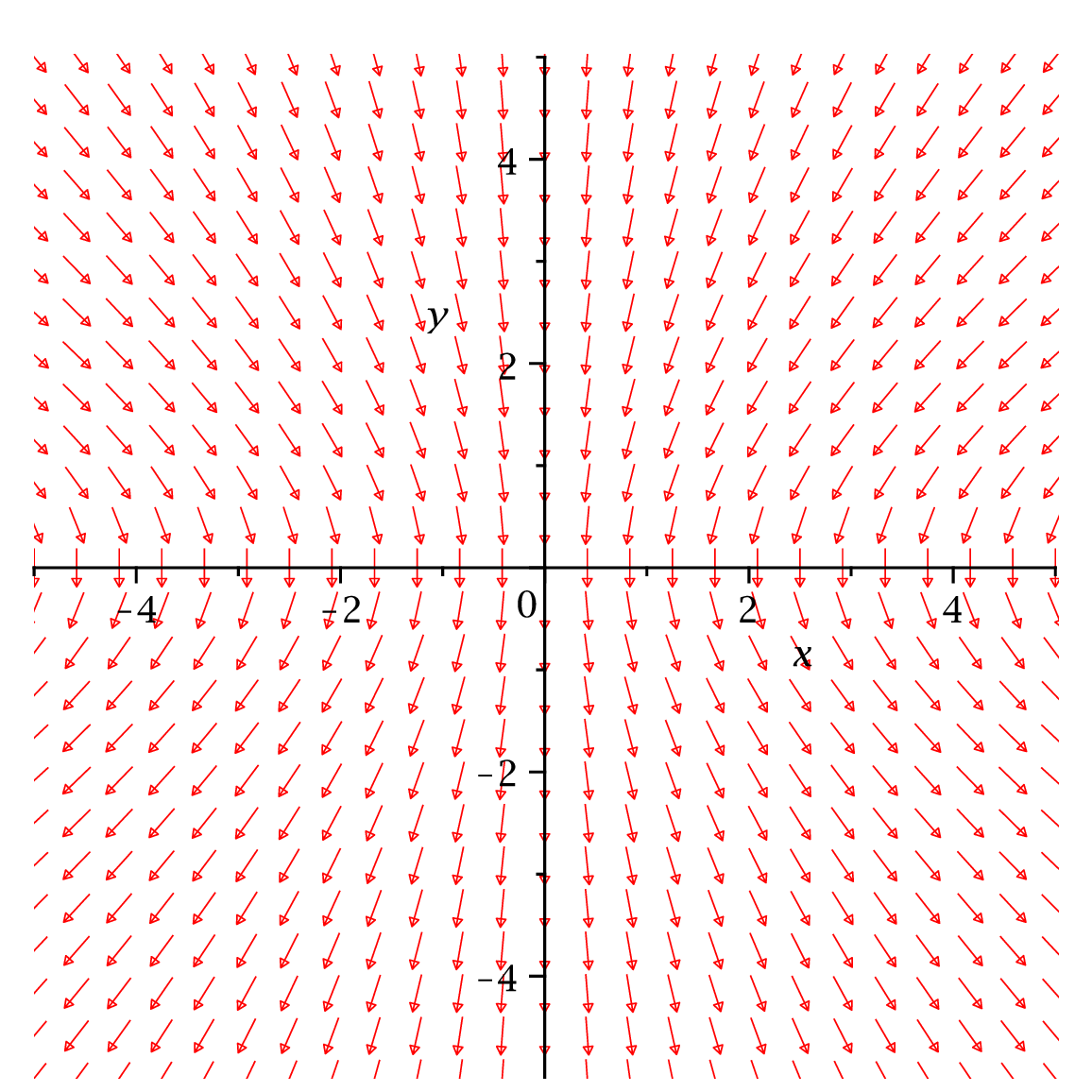}
    \caption{$n=1$.}
    \label{fig_vacB_AdSgrowing1}
  \end{subfigure}
  \begin{subfigure}[b]{0.49\textwidth}
    \centering
    \includegraphics[width=0.8\textwidth]{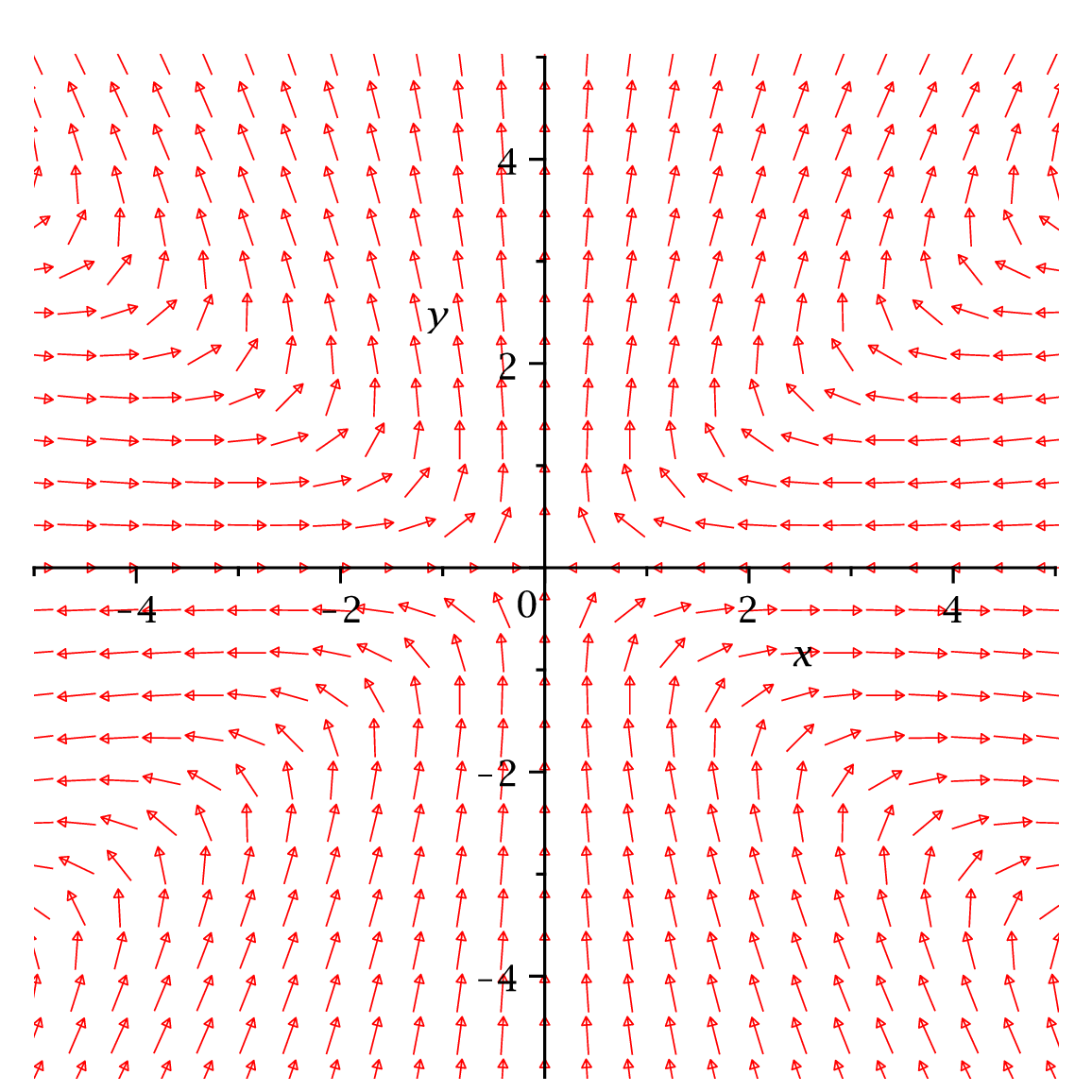}
    \caption{$n=2$.}
    \label{fig_vacB_AdSgrowing2}
  \end{subfigure}
  \begin{subfigure}[b]{0.49\textwidth}
    \centering
    \includegraphics[width=0.8\textwidth]{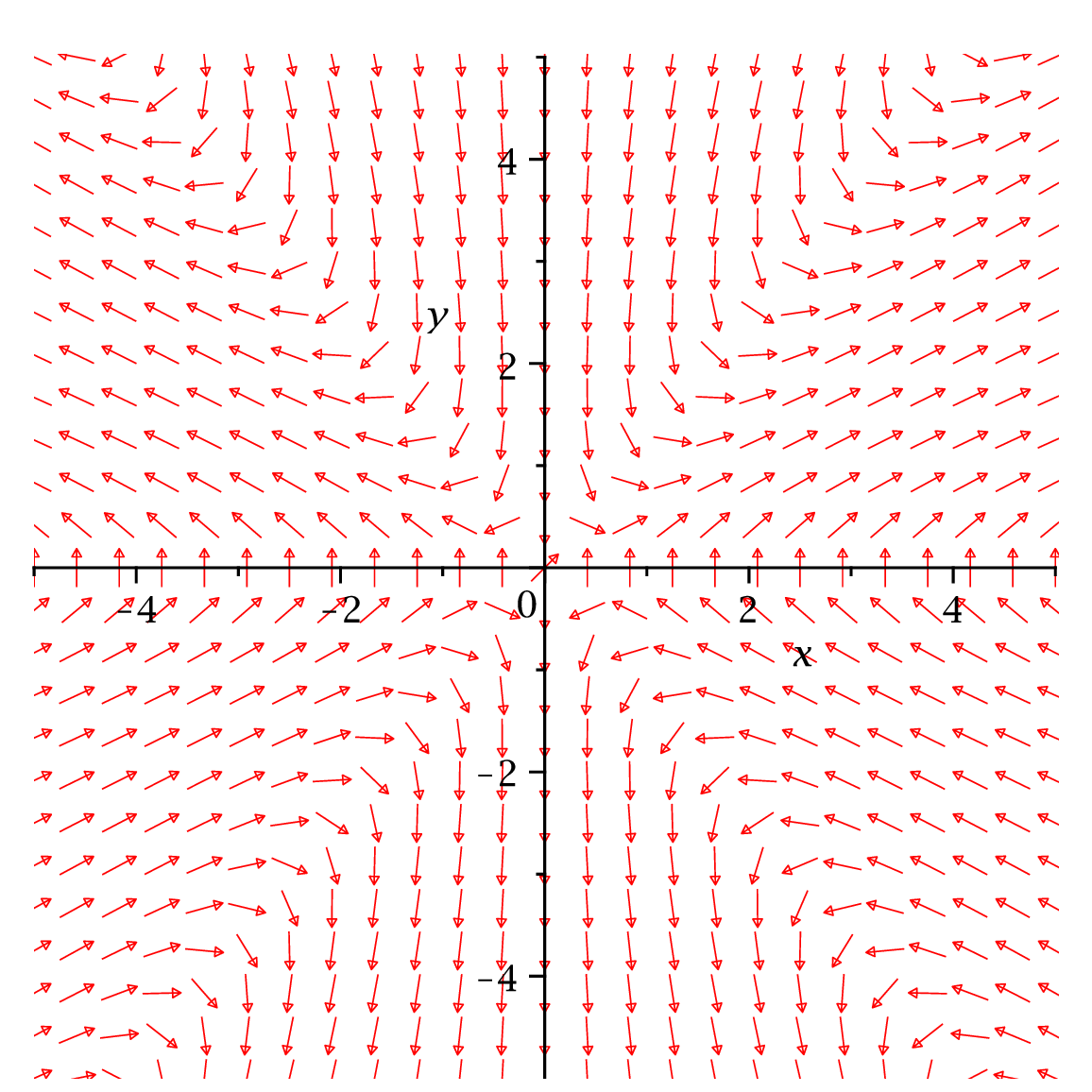}
    \caption{$n=3$.}
    \label{fig_vacB_AdSgrowing3}
  \end{subfigure}
  \begin{subfigure}[b]{0.49\textwidth}
    \centering
    \includegraphics[width=0.8\textwidth]{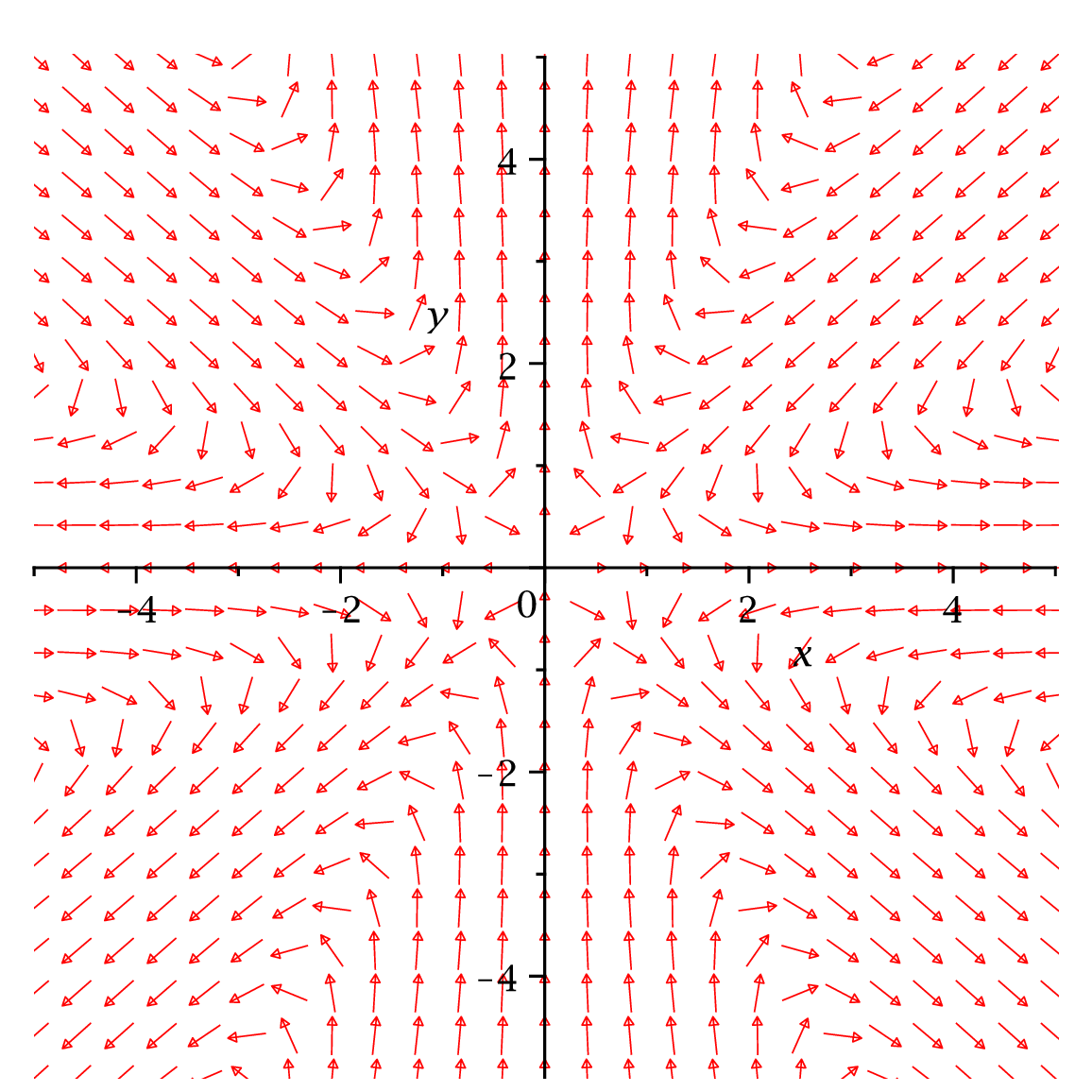}
    \caption{$n=4$.}
    \label{fig_vacB_AdSgrowing4}
  \end{subfigure}
  \caption{Plots of the spatial magnetic field configuration of the regular solutions \Eqref{eq_vacB_AdSgrowingsoln}. The axes are given in units of $L$.}
  \label{fig_vacB_AdSgrowing}
\end{figure}
The expressions for the energy density $\rho$ are somewhat complicated expressions involving $\tan^{-1}\brac{r/L}$ and trigonometric functions, so it is perhaps too cumbersome and not illuminating to be shown here. Nevertheless, these regular solutions do give a finite total energy over the whole spacetime, where
\begin{align}
 \mathcal{E}=\int_0^{2\pi}\dif\phi\int_0^\infty\dif r\int_0^\pi\dif\theta\;r^2\sin\theta\,\varrho.
\end{align}
For the four solutions in \Eqref{eq_vacB_AdSgrowingsoln}, their corresponding total energies are
\begin{align}
 n=1:\quad \mathcal{E}&=\frac{2\pi^2 b^2}{3\mu_0L},\nonumber\\
 n=2:\quad \mathcal{E}&=\frac{24\pi^2b^2}{5\mu_0L} ,\nonumber\\
 n=3:\quad\mathcal{E}&=\frac{3\pi^2 b^2}{7\mu_0L},\nonumber\\
 n=4:\quad\mathcal{E}&=\frac{256\pi^2b^2}{2205\mu_0L}.
\end{align}
The electric analogues of these regular solutions were studied in Ref.~\cite{Herdeiro:2015vaa}.

\subsection{Gauge potentials from Killing vectors} \label{sec_Maxwell_Killing}

In Ref.~\cite{Wald:1974np} Wald showed that a Killing vector of a Ricci-flat spacetime also solves the (source-free) Maxwell equation. Thus, a test electromagnetic field can be added to a Ricci-flat background by taking the gauge potential to be proportional to a Killing vector, giving another approach without having to perform separation of variables as was done in the previous subsections.

Here, we wish to take a similar approach of using Killing vectors. But in this present case, our (A)dS spacetimes are not Ricci-flat. We will presently see that the Maxwell equation requires the presence of a current source. To start, let $\xi^\mu$ be a Killing vector in a spacetime with metric $g_{\mu\nu}$ satisfying Killing's equation $\nabla_\mu\xi_\nu+\nabla_\nu\xi_\mu=0$. It can be shown\footnote{See, for instance, the paragraph following Eq.~(A13) of \cite{Geroch:1971} where Geroch explained how $\nabla_\mu\nabla_\nu\xi_\sigma=R_{\lambda\mu\nu\sigma}\xi^\lambda$ is obtained. Tracing over $\mu$ and $\nu$ then gives \Eqref{eq_Laplacian_xi1}.} that the following identity holds
\begin{align}
 \nabla^2\xi^\mu=-\xi^\lambda {R_\lambda}^\mu. \label{eq_Laplacian_xi1}
\end{align}
For (A)dS spacetimes we are considering, $R_{\mu\nu}=\Lambda g_{\mu\nu}$ so Eq.~\Eqref{eq_Laplacian_xi1} is
\begin{align}
 \nabla^2\xi^\mu=-\Lambda\xi^\mu, \label{eq_Laplacian_xi2}
\end{align}
which is a Helmholtz-type equation for $\xi^\mu$.

If the gauge potential $A_\mu$ is proportional to the Killing vector, the Killing's equation then implies $\nabla_\mu A_\nu=-\nabla_\nu A_\mu$. Therefore the Faraday tensor is $F_{\mu\nu}=2\nabla_\mu A_\nu$. Putting this into the Maxwell equation \Eqref{eq_ME_units} gives
\begin{align}
 \nabla^2A^\mu=-\frac{\mu_0}{2}J^\mu. \label{eq_Laplacian_A}
\end{align}
 With $A^\mu\propto \xi^\mu$ and comparing Eq.~\Eqref{eq_Laplacian_A} to \Eqref{eq_Laplacian_xi2}, we have
\begin{align}
 J^\mu=\frac{2\Lambda}{\mu_0}A^\mu.
\end{align}
Therefore if $\Lambda\neq0$, a Killing vector can be used as a solution to Maxwell's equation provided that the same vector is proportional to the current source.

For spacetimes of the form \Eqref{eq_PureBackground}, there is an axial Killing vector $ \frac{b}{2}\partial_\phi$, where $b$ is an arbitrary constant defined with a factor of half for later convenience. Lowering the index to get a one-form, it can be readily verified that the gauge potential
\begin{align}
 A=\half br^2\sin^2\theta\,\dif\phi \label{eq_KillingA}
\end{align}
is a solution to Maxwell's equation with current $J=\frac{\Lambda b}{\mu_0}\partial_\phi$. The flux through an area bounded by a circle $C$ of constant $r=R$ lying on the plane $\theta=\frac{\pi}{2}$ is
\begin{align}
 \mbox{\textsf{flux}}=\oint_CA_\mu\dif x^\mu=\pi r^2 b,
\end{align}
which is reminiscent of the elementary expression of $\brac{\mbox{\textsf{area}}}\times \brac{\mbox{\textsf{field strength}}}$ for a flux.

\begin{figure}
 \begin{subfigure}[b]{0.49\textwidth}
    \centering
    \includegraphics[width=0.8\textwidth]{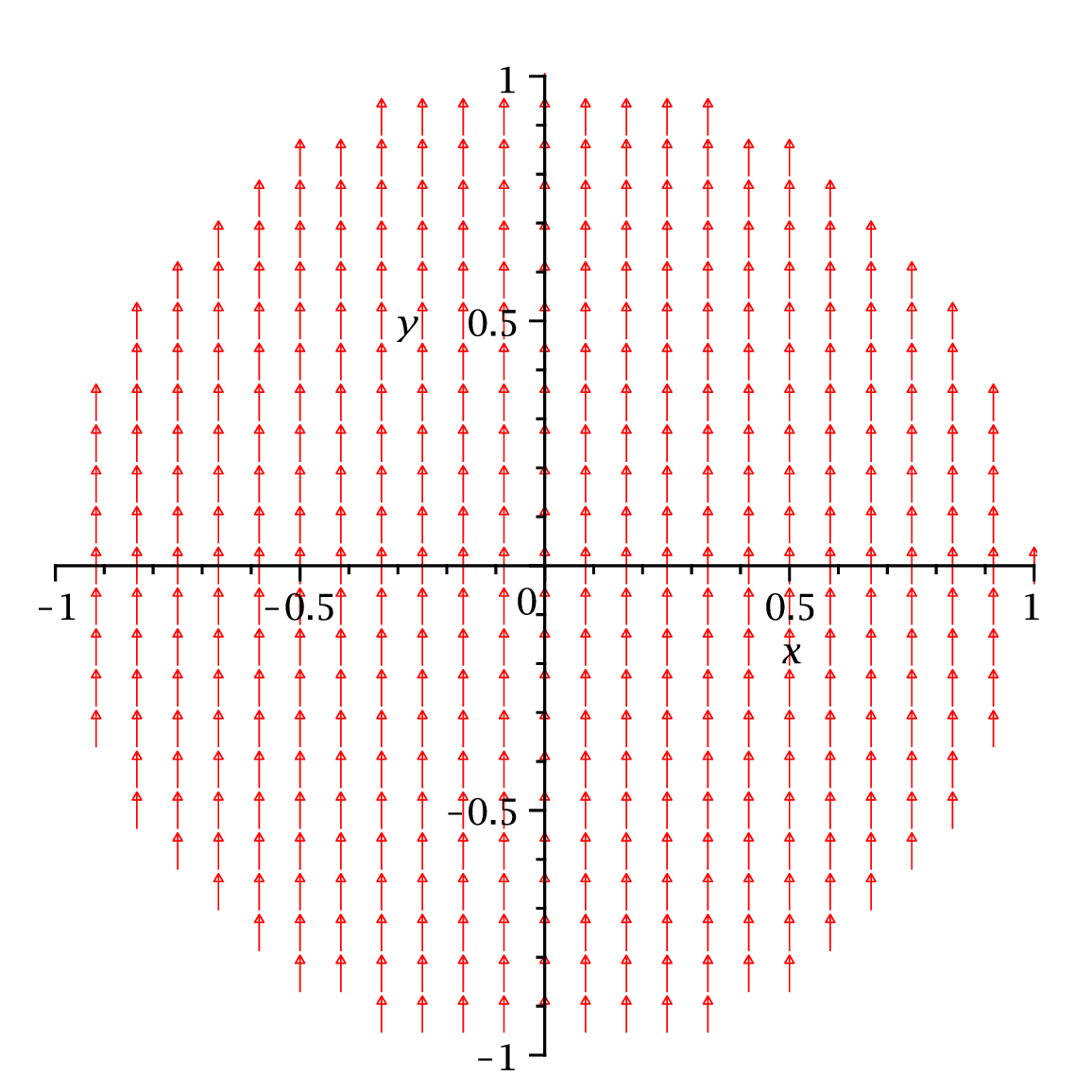}
    \caption{$\Lambda=3/\ell^2$ (dS).}
    \label{fig_vacB_dSKilling}
  \end{subfigure}
  \begin{subfigure}[b]{0.49\textwidth}
    \centering
    \includegraphics[width=0.8\textwidth]{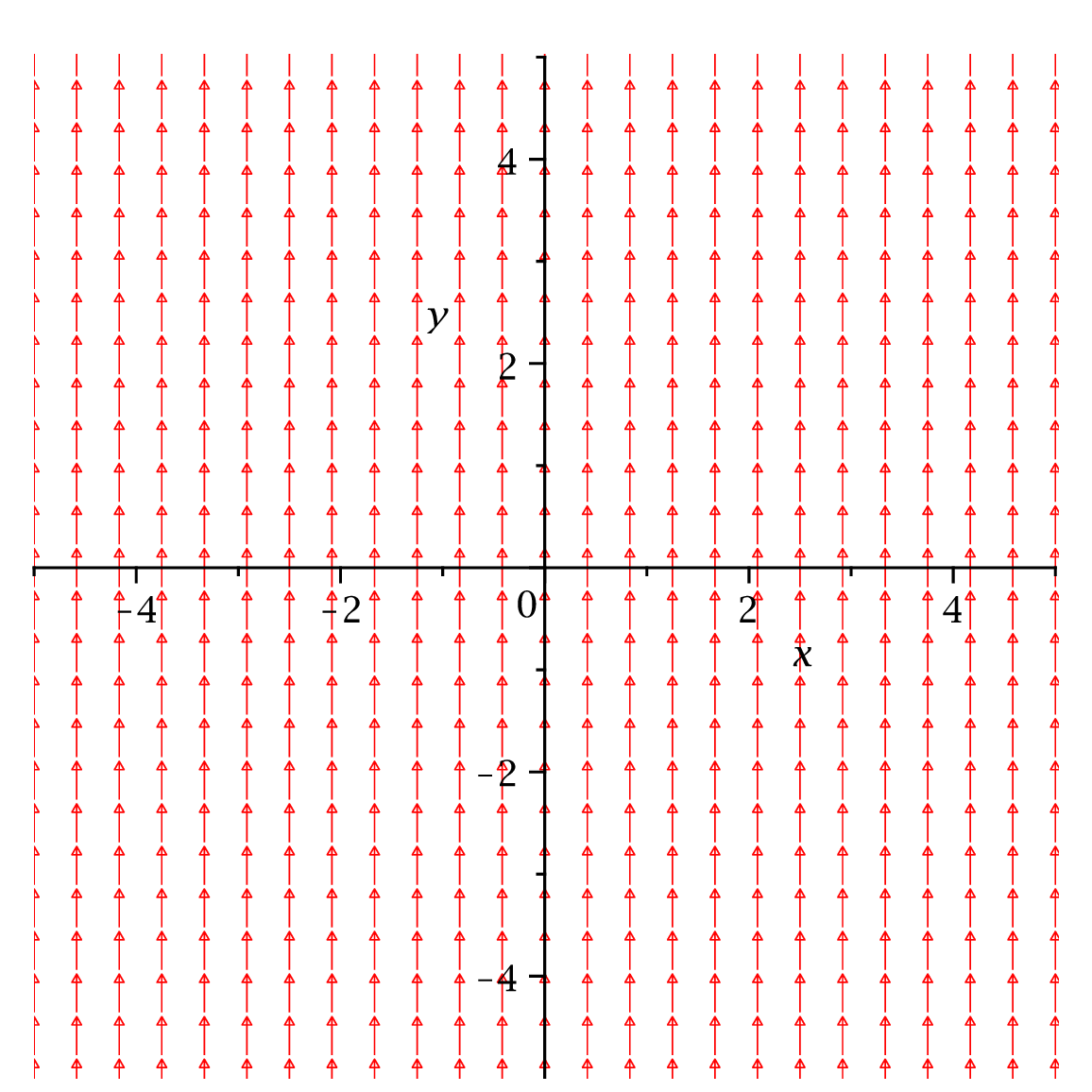}
    \caption{$\Lambda=-3/L^2$ (AdS).}
    \label{fig_vacB_AdSKilling}
  \end{subfigure}
  \caption{Plots of the magnetic field configuration arising from the potential given in \Eqref{eq_KillingA} for (a) $\Lambda>0$ and (b) $\Lambda<0.$}
  \label{fig_vacB_Killing}
\end{figure}

Using Eq.~\Eqref{eq_Bcomponents}, we see that the magnetic field configuration points along the $z$-direction, as shown in Fig.~\ref{fig_vacB_Killing}. The energy density is obtained using Eq.~\Eqref{eq_WeakVac_rho}, giving
\begin{align}
 \varrho=\frac{b^2}{6}\brac{3-\Lambda r^2\sin^2\theta}. \label{eq_WeakKilling_rho}
\end{align}
In particular it is regular at the origin, and is always positive for both dS ($\Lambda>0$) and AdS ($\Lambda<0$) cases. In the dS case, the density decreases with $r$ to the value $\frac{b^2}{2}\cos^2\theta$ at the cosmological  horizon. In the AdS case, the density grows quadratically with $r$. In this case as $r$ grows arbitrarily large, so does $\varrho$ and eventually the weak field description ceases to be valid.

\section{Magnetisation in the Einstein--Maxwell--fluid model} \label{sec_eom}

In this section we move beyond the weak field regime and seek magnetic fields that magnetise the (A)dS spacetime, while taking its gravitational backreaction into account. Therefore we do not expect the resulting spacetime to be perfectly (A)dS anymore, but some distortion thereof. We will find this solution by solving the Einstein--Maxwell equations in the presence of an anisotropic fluid.

\subsection{Equations of motion and the magnetisation procedure}

In the following, it will be convenient to work in geometric units. This amounts to absorbing the physical constants into the expressions of the matter sources by
\begin{align}
 \sqrt{\frac{4\pi G}{\mu_0}}F_{\mu\nu}\rightarrow F_{\mu\nu},\quad GT_{\mu\nu}\rightarrow T_{\mu\nu},\quad \mu_0J^\mu\rightarrow J^\mu. \label{eq_ToGeometricUnits}
\end{align}
It then follows that the gauge potential is consequently redefined as $\sqrt{\frac{4\pi G}{\mu_0}}A_\mu\rightarrow A_\mu$. The equations of motion are now expressed in the form
\begin{subequations}\label{EinsteinMaxwellEqns}
\begin{align}
 R_{\mu\nu}&=2F_{\mu\lambda}{F_\nu}^\lambda-\frac{1}{2}F^2g_{\mu\nu}+8\pi\brac{T_{\mu\nu}-\frac{1}{2}{T^\lambda}_\lambda g_{\mu\nu}},\\
 \nabla_\mu F^{\mu\nu}&=-J^\nu.
\end{align}
\end{subequations}
We shall take the following ansatz for the metric and gauge potential
\begin{subequations}\label{ansatz_metric_potential}
\begin{align}
 \dif s^2&=\expo{2U}\dif\sigma^2+\expo{2U}\bar{g}_{ab}\dif x^a\dif x^b,\label{ansatz_metric}\\
  A&=\chi\dif\sigma,\label{ansatz_potential}
\end{align}
\end{subequations}
where $U$ and $\chi$ are functions independent of $\sigma$, and $\bar{g}_{ab}$ is a three dimensional metric of Lorentzian signature. Here our notation for coordinate indices are $x^\mu=(x^a,\sigma)$, where Greek indices $\mu,\nu,\lambda,\ldots$ covers all four coordinates, and lowercase Latin indices $a,b,c,\ldots$ represent the other three coordinates $x^a$.

For the matter sources we shall take the anisotropic fluid
\begin{align}
 T_{\mu\nu}=(\rho+p)u_\mu u_\nu+pg_{\mu\nu}+(p_\sigma-p)n_\mu n_\nu, \label{ansatz_fluid}
\end{align}
where $\rho$ is the mass density of the fluid, $p$ and $p_\sigma$ are the anisotropic pressures, $u^\mu$ is the $4$-velocity of the fluid and $n^\mu$ is a unit spacelike vector in the direction of anisotropy. Isotropy is recovered when $p_\sigma=p$. We will take $u^\mu$ and $n^\mu$ to be
\begin{align}
 u^\mu=\brac{\expo{U}v^a,0},\quad n^\mu=\brac{0,0,0,\expo{-U}},
\end{align}
where $v^a$ is a normalised timelike $3$-velocity within $\bar{g}_{ab}$ such that $\bar{g}_{ab}v^av^b=-1$. The normalisations $g_{\mu\nu}u^\mu u^\nu=-1$ and $g_{\mu\nu}n^\mu n^\nu=+1$ can be directly verified. The electromagnetic $4$-current is taken to be
\begin{align}
 J^\mu=\brac{0,0,0,\mathcal{J}},
\end{align}
for some current component $\mathcal{J}$.

With the setup above, the equations of motion become
\begin{subequations} \label{EinsteinMaxwellFluid_eom}
\begin{align}
 \bar{R}_{ab}=2\big(\bar{\nabla}_aU\bar{\nabla}_bU&+\expo{-2U}\bar{\nabla}_a\chi\bar{\nabla}_b\chi\big)+8\pi\sbrac{(\rho+p)v_av_b+\brac{\rho-p}\bar{g}_{ab}}\expo{-2U},\\
 \bar{\nabla}^2U+\expo{-2U}\brac{\bar{\nabla}\chi}^2&=-4\pi\brac{\rho-2p+p_\sigma}\expo{-2U},\\
 \bar{\nabla}\cdot\brac{\expo{-2U}\bar{\nabla}\chi}&=-\mathcal{J}\expo{-2U}.
\end{align}
\end{subequations}
Let us briefly recall how Ricci-flat solutions were magnetised using Harrison transformations. In the Ricci-flat case, $R_{\mu\nu}=0$ and the current-carrying fluid is absent, we have $\rho=p=p_0=\mathcal{J}=0$ and the equations of motion reduce to
\begin{align}
 \bar{R}_{ab}=2\big(\bar{\nabla}_aU\bar{\nabla}_bU&+\expo{-2U}\bar{\nabla}_a\chi\bar{\nabla}_b\chi\big),\nonumber\\
 \bar{\nabla}^2U+\expo{-2U}\brac{\bar{\nabla}\chi}^2&=0,\nonumber\\
 \bar{\nabla}\cdot\brac{\expo{-2U}\bar{\nabla}\chi}&=0.
\end{align}
This is just the Einstein--Maxwell equations under the ansatz \Eqref{ansatz_metric_potential}, which is well known to be invariant under the Harrison transformation
\begin{align}
  U\rightarrow U'&=U-\ln H,\quad \chi\rightarrow\chi'=H^{-1}\sbrac{\chi+b\brac{\expo{2U}+\chi^2}},\nonumber\\
 H&=\brac{1+b\chi}^2+ b^2\expo{2U}. \label{HarrisonTransform_ori}
\end{align}
This is the usual way one derives magnetised spacetimes in the absence of a cosmological constant\cite{Dowker:1993bt,Ortaggio:2004kr}. Let us briefly review how this works. Suppose we have a vaccum solution, which in terms of the ansatz \Eqref{ansatz_metric_potential} corresponds to $(U,\chi=0)$ so $U$ is a function which solves
\begin{align}
 \bar{R}_{ab}=2\bar{\nabla}_aU\bar{\nabla}_bU,\quad\bar{\nabla}^2U=0. \label{EinsteinMaxwell_seed_eom}
\end{align}
One then seeks a new solution $(U',\chi')$, which would be required to satisfy the equations
\begin{align}
 \bar{R}_{ab}=2\big(\bar{\nabla}_aU'\bar{\nabla}_bU'&+\expo{-2U'}\bar{\nabla}_a\chi'\bar{\nabla}_b\chi'\big),\nonumber\\
 \bar{\nabla}^2U'+\expo{-2U'}\brac{\bar{\nabla}\chi'}^2&=0,\nonumber\\
 \bar{\nabla}\cdot\brac{\expo{-2U'}\bar{\nabla}\chi'}&=0.\label{EinsteinMaxwell_transformed_eom}
\end{align}
If $U'$ and $\chi'$ are generated using Eq.~\Eqref{HarrisonTransform_ori}, we have
\begin{align*}
 U'=U-\ln H,\quad \chi'=bH^{-1}\expo{2U},\quad H=1+b^2\expo{2U}
\end{align*}
for some constant $b$. Then we can see that $(U',\chi')$ satisfies \Eqref{EinsteinMaxwell_transformed_eom} because the seed $(U,\chi=0)$ satisfies \Eqref{EinsteinMaxwell_seed_eom}.

Now suppose that we wish to magnetise the (A)dS spacetime. Then the seed would be a spacetime metric whose Ricci tensor obeys
\begin{align}
 R_{\mu\nu}=\Lambda g_{\mu\nu}, \label{EinsteinSpace}
\end{align}
where $\Lambda$ is the cosmological constant. A possible framework to consider is by simply including a cosmological constant to the Einstein--Maxwell model. That is, by having $\mathcal{L}_{\mathrm{m}}=-2\Lambda$ in the action \Eqref{action}. However, upon using the ansatz \Eqref{ansatz_metric_potential}, the equations of motion are
\begin{align}
 \bar{R}_{ab}=2\big(\bar{\nabla}_aU\bar{\nabla}_bU&+\expo{-2U}\bar{\nabla}_a\chi\bar{\nabla}_b\chi\big)+2\Lambda\expo{-2U},\nonumber\\
 \bar{\nabla}^2U+\expo{-2U}\brac{\bar{\nabla}\chi}^2&=-\Lambda\expo{-2U},\nonumber\\
 \bar{\nabla}\cdot\brac{\expo{-2U}\bar{\nabla}\chi}&=0.
\end{align}
The presence of $\Lambda$ breaks the symmetry of the Harrison transform and the equations of motion are no longer invariant under \Eqref{HarrisonTransform_ori}.

On the other hand, if we instead regard the (A)dS spacetime as being sourced by a fluid with the equation of state
\begin{align*}
 \rho=\frac{\Lambda}{8\pi}=-p=-p_\sigma.
\end{align*}
we still can apply a Harrison-like transformation, supplemented by appropriate transformations of $\rho$, $p$, $p_\sigma$, and $\mathcal{J}$ that will still solve the equations of motion. In previous literature, this method has been successfully applied to magnetise stars sourced by anisotropic fluids \cite{Stelea:2018cgm} and black holes in the presence of quitessential matter \cite{Lungu:2024iob}.

With these considerations, we now turn to Eq.~\Eqref{EinsteinMaxwellFluid_eom} and ask whether one could magnetise a dS or AdS seed that initially is free of electromagnetic fields. Suppose that we have a seed solution that is described by functions $(U,\chi=0,\rho,p,p_\sigma,\mathcal{J}=0)$. Then the non-zero functions satisfies
\begin{align}
 \bar{R}_{ab}&=2\bar{\nabla}_aU\bar{\nabla}_bU+8\pi\sbrac{(\rho+p)v_av_b+\brac{\rho-p}\bar{g}_{ab}}\expo{-2U},\nonumber\\
 \bar{\nabla}^2U&=-4\pi\brac{\rho-2p+p_\sigma}\expo{-2U}.\label{EinsteinMaxwellFluid_seed_eom}
\end{align}
We seek a new solution described by the functions $(U',\chi',\rho',p',p_\sigma',\mathcal{J}')$, which means they must obey \Eqref{EinsteinMaxwellFluid_eom}. We construct a solution by taking
\begin{align}
 U'&=U-\ln H,\quad\chi'=\frac{c\expo{2U}}{H},\quad H=1+c^2\expo{2U},\nonumber\\
 \rho'&=\rho H^{-2},\quad p'=pH^{-2},\quad p_\sigma'=p_\sigma H^{-2}-2c^2H^{-3/2}\expo{2U}\sbrac{{\rho-2p}+p_\sigma},\nonumber\\
 \mathcal{J}'&=8\pi c\brac{\rho-2p+p_\sigma}H^{-2}, \label{HarrisonFluid_transform}
\end{align}
for some constant $c$. One can directly verify that $(U',\chi',\rho',p',p_\sigma',\mathcal{J}')$ constructed as above will solve \Eqref{EinsteinMaxwellFluid_eom} due to the fact that the seed $(U,\chi=0,\rho,p,p_\sigma,\mathcal{J}=0)$ satisfies \Eqref{EinsteinMaxwellFluid_seed_eom}.

\subsection{Magnetising with an (A)dS seed}

We will now apply the magnetisation procedure to the (A)dS spacetime, where the metric was given in Eq.~\Eqref{eq_PureBackground}. For convenience, we write here again:
\begin{subequations} \label{metric_AdS}
 \begin{align}
  \dif s^2&=-f(r)\dif t^2+f(r)^{-1}\dif r^2+r^2\brac{\dif\theta^2+\sin^2\theta\,\dif\phi^2},\label{metric_AdS_ds2}\\
  f(r)&=1-\frac{\Lambda}{3}r^2. \label{metric_AdS_f}
 \end{align}
\end{subequations}
The Ricci tensor of this metric satisfies \Eqref{EinsteinSpace}. In terms of the Einstein--Maxwell--fluid model, this is a solution in which the fluid has the equation of state
\begin{align}
 \rho=\frac{\Lambda}{8\pi}=-p=-p_\sigma, \label{seed_fluid}
\end{align}
with vanishing gauge potential $\chi=0$.

To apply the Harrison-type magnetising procedure to this solution, we cast \Eqref{metric_AdS} into the form of the ansatz \Eqref{ansatz_metric_potential},
\begin{align*}
 \dif s^2&=\brac{\frac{r}{L}\sin\theta}^2\dif\sigma^2+\brac{\frac{r}{L}\sin\theta}^{-2}\brac{\frac{r}{L}\sin\theta}^{2}\brac{-f\dif t^2+\frac{\dif r^2}{f}+r^2\dif\theta^2},
\end{align*}
where $\sigma=L\phi$, and $L$ is an arbitrary length parameter to ensure dimensional consistency. (That is, $U$ is dimensionless and $\sigma$ is a coordinate with the dimension of length.) We thus identify
\begin{align}
 \bar{g}_{ab}\dif x^a\dif x^b&=\brac{\frac{r}{L}\sin\theta}^{2}\brac{-f\dif t^2+\frac{\dif r^2}{f}+r^2\dif\theta^2},\quad U=\ln\brac{\frac{r}{L}\sin\theta}. \label{seed_U}
\end{align}
Upon applying the transformation \Eqref{HarrisonFluid_transform} and dropping the primes on the new solution, the result is
\begin{subequations}\label{4d_soln}
\begin{align}
  \dif s^2&=H^2\brac{-f\dif t^2+\frac{\dif r^2}{f}+r^2\dif\theta^2}+\frac{r^2\sin^2\theta}{H^2}\dif\phi^2,\\
  H&=1+\frac{1}{4}B^2r^2\sin^2\theta, \quad f=1-\frac{\Lambda}{3}r^2,\label{4d_metric}\\
  A&=\frac{Br^2\sin^2\theta}{2H}\dif\phi,\quad J=-\frac{B\Lambda}{H^2}\partial_\phi,\\
  \rho&=\frac{\Lambda}{8\pi H^2}=-p_1=-p_2,\quad p_\sigma=-\rho-\frac{B^2\Lambda r^2\sin^2\theta}{4\pi H^3}.\label{4d_rho_p}
\end{align}
\end{subequations}
where we have let $c=\half B$. The solution \Eqref{4d_soln} can be directly verified to satisfy the Einstein--Maxwell--fluid equations \Eqref{EinsteinMaxwellEqns}.

This solution is a generalisation of the Killing vector gauge potential solution \Eqref{eq_KillingA}, where backreaction to the spacetime is taken into account. To see this, we restore the standard units on the magnetic field parameter by
\begin{align}
 B=\sqrt{\frac{4\pi G}{\mu_0}}b, \label{eq_RestoreSI}
\end{align}
where $b$ is the magnetic field given in appropriate standard units. (Such as $\mathrm{T}/c^2$.) In the weak field limit, the magnetic field is sufficiently weak so as not to give gravitational effect. This means $B^2=4\pi Gb^2/\mu_0$ is negligible so $H^2\simeq 1$ and \Eqref{4d_metric} reduces to \Eqref{metric_AdS_ds2}, whereas the gauge potential reduces to the weak field solution \Eqref{eq_KillingA}.

\section{Geometrical and physical properties} \label{sec_properties}

\subsection*{The cosmological horizon}
In the dS case ($\Lambda>0$), the spacetime has a cosmological horizon at $r=\ell$, where $\ell=\sqrt{3/\Lambda}$. The spacetime has a time-like Killing vector $\xi^\mu=(C,0,0,0)$, where $C$ is an arbitrary normalisation constant. It can be easily verified to satisfy the Killing equation $\nabla_\mu\xi_\nu+\nabla_\nu\xi_\mu=0$. Its norm is
\begin{align}
 \xi^\mu\xi_\mu=-C^2H^2f,
\end{align}
which vanishes at $r=\ell=\sqrt{3/\Lambda}$, indicating the presence of a horizon.

The horizon area can be computed at constant $t$ and $r=\ell$, giving
\begin{align}
 \mathcal{A}=4\pi\ell^2,
\end{align}
which is the same as the usual dS horizon, and is unaffected by the presence of the magnetic field. The surface gravity is also unaffected by the magnetic field. The surface gravity at the horizon is evaluated using
\begin{align}
 \kappa=\lim_{r\rightarrow\ell}\sqrt{-\half\nabla_\mu\xi_\nu\nabla^\mu\xi^\nu}=\frac{C}{\ell},
\end{align}
which is the same as the usual dS surface gravity and unaffected by the magnetic field.

\subsection*{Embedding diagrams}
Surfaces of constant $t$ and $r=a$ is
\begin{align}
 \dif s_a^2=H_a^2a^2\dif\theta^2+\frac{a^2\sin^2\theta}{H_a^2}\dif\phi^2, \label{const_tr}
\end{align}
where $H_a=1+\frac{1}{4}B^2a^2\sin^2\theta$. Note that this sector of the metric is independent of $f$ and therefore the geometry will be same for both dS and AdS cases.

We attempt to visualise the geometry of \Eqref{const_tr} by embeddding it into a Euclidean 3-space
\begin{align}
 \dif s^2_{\mathbb{R}^3}=\dif R^2+R^2\dif\Phi^2+\dif Z^2.
\end{align}
To find the embedding surface, we parametrise the surface with $R=R(\theta)$, $Z=Z(\theta)$, and $\Phi=\phi$.
Then the induced metric is $\dif s^2=\brac{R'^2+Z'^2}\dif\theta^2+R^2\dif\phi^2$.
Comparing this with \Eqref{const_tr} we find,
\begin{align*}
 R=\frac{a\sin\theta}{H_a},\quad Z'=\pm\sqrt{a^2H_a^2-R'^2}.
\end{align*}
We integrate the first order equation for $Z$ and then perform a surface of revolution about the $Z$ axis.

\begin{figure}[htb!]
 \centering
 \includegraphics{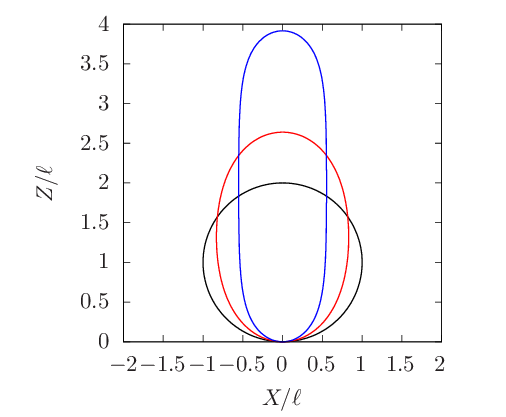}
 \caption{Embedding diagram of dS horizon geometry for $B\ell=0$, $0.9$, and $1.8$.}
 \label{fig_embed}
\end{figure}

The geometry of the dS horizon $a=\ell$ is shown in  Fig.~\ref{fig_embed}. For $B=0$, we get a round sphere, as expected. For increasing $B>0$, the magnetic field appears to distort the horizon by stretching it along the $Z$-direction.

\subsection*{Curvature invariants}
With the metric \Eqref{4d_metric}, we can compute its various curvature invariants, $R$, $R_{\mu\nu}R^{\mu\nu}$, and $R_{\mu\nu\rho\sigma}R^{\mu\nu\rho\sigma}$. Its full general expressions are somewhat cumbersome to display here. But it can be checked that no curvature singularities are present. For specific values of $\theta$, the Kretschmann invariant can be displayed as
\begin{align}
 \left.R_{\mu\nu\rho\sigma}R^{\mu\nu\rho\sigma}\right|_{\theta=0}&=\frac{8}{3}\Lambda^2+20B^4,\nonumber\\
 \left.R_{\mu\nu\rho\sigma}R^{\mu\nu\rho\sigma}\right|_{\theta=\pi/4}&=\frac{32768}{9(8+B^2r^2)^8}\big(-13824B^6r^2+864B^8r^4+92160B^4+12288\Lambda^2\nonumber\\
 &\hspace{1cm}-39936\Lambda r^2B^4+63\Lambda^2r^8B^8-288\Lambda r^6B^8+528\Lambda^2 r^6B^6+3456\Lambda r^4B^6\nonumber\\
 &\hspace{1cm}+9216 \Lambda^2r^2B^2+8320\Lambda^2r^4B^4\big),\nonumber\\
 \left.R_{\mu\nu\rho\sigma}R^{\mu\nu\rho\sigma}\right|_{\theta=\pi/2}&=\frac{2048}{9(4+B^2r^2)^8}\big(-1728B^6r^2+216B^8r^4+5760B^4+768\Lambda^2-4992\Lambda r^2B^4\nonumber\\
 &\hspace{1cm}+63\Lambda^2r^8B^8-144\Lambda r^6B^8+264\Lambda^2r^6B^6+864\Lambda r^4B^6\nonumber\\
 &\hspace{1cm}+1152\Lambda^2r^2B^2+2080\Lambda^2r^4B^4\big).
\end{align}

\subsection*{Energy of the matter fields}
From Eq.~\Eqref{4d_rho_p} we see that the energy density and pressures of the fluid obey
\begin{align}
 \rho+p_1&=0,\quad \rho+p_2=0,\quad\rho+p_\sigma=-\frac{B^2\Lambda r^2\sin^2\theta}{4\pi H^3}.
\end{align}
Therefore, the magnetised AdS solution ($\Lambda<0$) obeys the null energy condition, but the dS solution ($\Lambda>0$) violates it. In other words, the magnetised dS spacetime as constructed above requires the presence of exotic matter.

Turning to the energy density the magnetic field itself, we use Eq.~\Eqref{eq_EMrho_def} for the present solution, the energy density of the magnetic field seen by a static observer with four-velocity $u^\mu$ is\footnote{Note that for the present metric, the four-velocity of a static observer is $u^\mu=\brac{1/f^{1/2}H,0,0,0}$.}
\begin{align}
 \varrho=\frac{B^2\brac{3-\Lambda r^2\sin^2\theta}}{6H^4}.
\end{align}
Note that in the weak field limit, $H\simeq1$ and we recover the energy density \Eqref{eq_WeakKilling_rho} of the weak field solution.

\subsection*{Flux}
Let us consider a region $D$ bounded by a circle $C$ of radius $r=R$ centred at the origin and lying in the equatorial plane $\theta=\frac{\pi}{2}$. The flux passing through $D$ is
\begin{align}
 \Phi=\int_D F=\oint_C A=\int_0^{2\pi}\frac{BR^2\sin^2\theta}{2\brac{1+\frac{1}{4}B^2R^2}}\dif\phi=\frac{\pi R^2 B}{1+\frac{1}{4}B^2R^2},
\end{align}
which is independent of $\Lambda$ and is therefore similar to the standard $\Lambda=0$ Melvin case. However for dS $(\Lambda>0)$, one may obtain the total flux within the static patch $0<r<\ell$, where $\ell=\sqrt{3/\Lambda}$,
\begin{align}
 \mbox{dS}:\quad \Phi_{\mathrm{tot}}=\frac{\pi B\ell^2}{1+\frac{1}{4}B^2\ell^2}.
\end{align}
For AdS $(\Lambda<0)$, the total flux is
\begin{align}
 \mbox{AdS}:\quad \Phi_{\mathrm{tot}}=\lim_{R\rightarrow\infty}\Phi=\frac{4\pi}{B},
\end{align}
which is inversely proportional to $B$, similar to the planar Melvin-AdS solution studied by Kastor and Traschen \cite{Kastor:2020wsm}.

\section{Geodesics} \label{sec_geod}

One intuitive way to understand the gravitational field of the spacetime is to study the trajectories of test particles within it. So in this section we shall briefly study null and time-like geodesics in the spacetime. For concreteness, we restrict our attention to equatorial geodesics in the plane $\theta=\frac{\pi}{2}$ where the motion can be described in terms of an effective potential that depends only on the radial coordinate. Then, circular orbits and domains of motion can be studied systematically. In fact, as we will see below it suffices to study the numerator of the effective potential, which is a polynomial in $x=r^2$.

\subsection{Effective potential for equatorial motion}

The geodesics describe the trajectory of test particles as parametrised curves  $x^\mu(\tau)$ where $\tau$ is an appropriate affine parametrisation. The four-velocity is $\dot{x}^\mu=\frac{\dif x^\mu}{\dif\tau}$, where over-dots denote derivatives with respect to $\tau$. The norm-squared of the four-velocity is
\begin{align}
 \epsilon&=g_{\mu\nu}\dot{x}^\mu\dot{x}^\nu=H^2\brac{-f\dot{t}^2+\frac{\dot{r}^2}{f}+r^2\dot{\theta}^2}+\frac{r^2\sin^2\theta}{H^2}\dot{\phi}^2. \label{eq_4velocity_normalisation}
\end{align}
One can rescale the affine parameter such that the norm-squared is 
\begin{align}
\epsilon =
 \begin{cases}
  -1 & \text{time-like,}\\
  0 & \text{null.}
 \end{cases}
\end{align}

The trajectory is governed by the Lagrangian $\mathcal{L}=\half g_{\mu\nu}\dot{x}^\mu\dot{x}^\nu$. For the metric \Eqref{4d_soln} the Lagrangian is explicitly
\begin{align}
 \mathcal{L}=\half\sbrac{H^2\brac{-f\dot{t}^2+\frac{\dot{r}^2}{f}+r^2\dot{\theta}^2}+\frac{r^2\sin^2\theta}{H^2}\dot{\phi}^2}.
\end{align}
Since $t$ and $\phi$ are cyclic variables, we have the following conserved quantities
\begin{align}
 \frac{\partial\mathcal{L}}{\partial\dot{t}}=-H^2f\dot{t}=-E,\quad\frac{\partial\mathcal{L}}{\partial\dot{\phi}}=\frac{r^2\sin^2\theta}{H^2}\dot{\phi}=L,\label{eq_FirstIntegrals}
\end{align}
where we will regard $E$ and $L$ to be the particle's conserved energy and angular momentum, respectively. In the following, we shall consider only future-directed trajectories so we take $E>0$. Furthermore the Lagrangian is invariant under the replacements
\begin{align}
 L\rightarrow-L,\quad B\rightarrow -B,
\end{align}
so we may take $L\geq 0$ and $B\geq0$ without loss of generality. Using Eq.~\Eqref{eq_FirstIntegrals} to express $\dot{t}$ and $\dot{\phi}$ in terms of $E$ and $L$, Eq.~\Eqref{eq_4velocity_normalisation} becomes
\begin{align}
 H^2\brac{\dot{r}^2+r^2f\dot{\theta}^2}-\frac{E^2}{H^2}+f\brac{\frac{H^2L^2}{r^2\sin^2\theta}-\epsilon}=0. \label{eq_FirstIntegral2}
\end{align}

For concreteness, we shall consider equatorial motion, where $\theta=\frac{\pi}{2}=\mathrm{constant}$. It can be verified that a constant $\theta=\frac{\pi}{2}$ satisfies the Euler--Lagrange equation for $\theta$. In this case then Eq.~\Eqref{eq_FirstIntegral2} becomes
\begin{align}
 H^2\dot{r}^2+U=0,\quad U=f\brac{\frac{H^2L^2}{r^2}-\epsilon}-\frac{E^2}{H^2},\label{eq_EffectiveEnergy}
\end{align}
where $U$ is the \emph{effective potential}. Since $H^2\dot{r}^2$ cannot be negative in Eq.~\Eqref{eq_EffectiveEnergy}, the domain of allowed motion must be such that $U<0$. Furthermore, \emph{circular orbits} are those where $r$ is constant along the trajectory. Such an orbit satisfies
\begin{align}
 U=U'=0, \label{eq_CircularOrbitCondition}
\end{align}
where primes denote derivatives with respect to $r$. From these facts we can glean some useful information about the time-like and null trajectories, as we will consider below in turn.

\subsection{Null geodesics}

For null geodesics we have $\epsilon=0$. In this case, it is convenient to write the effective potential as $U=E^2V$, where
\begin{align}
 V=f\frac{H^2}{r^2}\eta^2-\frac{1}{H^2}, \label{eq_NullGeod_EffectivePotential}
\end{align}
and we have defined $\eta=L/E$. Note that, with $U=E^2V$, the effective potential equation is $H^2\dot{r}+E^2V=0$. Since the four-velocity is null, there still remains a freedom in rescaling $\tau$ to remove $E^2$. So, for fixed $B$ and $\Lambda$, the family of null geodesics depend on the ratio $\eta$.

Note that in Eq.~\Eqref{eq_NullGeod_EffectivePotential}, $r$ always appear in even powers. Therefore it is convenient to introduce the parametrisation
\begin{align}\label{eq_xtor}
 x = r^2,
\end{align}
so that only non-negative $x$ are considered since the roots are described as $r=\sqrt{x}$. Then, Eq.~\Eqref{eq_NullGeod_EffectivePotential} becomes
\begin{subequations}
\begin{align}
 V&=\frac{P}{48x(4+B^2x)^2},\\
 P&=-\eta^2\Lambda x^5B^8+B^6\eta^2(3B^2-16\Lambda)x^4+48B^4\eta^2(B^2-2\Lambda)x^3+32B^2 \eta^2(9B^2-8\Lambda)x^2\nonumber\\
  &\hspace{2cm}+(768\eta^2B^2-768-256\eta^2\Lambda)x+768\eta^2. \label{eq_NullGeod_Pdef}
\end{align}
\end{subequations}
Since the denominator of $V$ is always positive, the condition for the domain of allowed motion, $U<0$, is now expressed as $P<0$. The non-negative roots of the degree-5 polynomial $P$ then defines the boundary of the allowed motion.

To determine the distribution of roots, we regard $P$ as a family of polynomials parametrised by $B$ and $\eta$, keeping $\Lambda$ as a fixed scale. Since we can consider positive $B$ and $\eta$ without loss of generality, and they always appear as even powers in $P$, we take the parameter space to be $(B^2,\eta^2)$, whose values then determine whether $P(x)$ has real, degenerate, or complex roots. The points where $P$ has degenerate roots are characterised by its vanishing discriminant, or equivalently, its \emph{$A$-discriminant} \cite{Lim:2021lju}. The $A$-discriminant is defined the condition
\begin{align}
 P(x)=P'(x)=0.\label{eq_NullGeod_PPp}
\end{align}
Note that this also defines circular orbit in which $r=\sqrt{x}$ is constant along the trajectory since it makes $\dot{r}=0$ in Eq.~\Eqref{eq_EffectiveEnergy}. Viewing Eq.~\Eqref{eq_NullGeod_PPp} as simultaneous equations for $B^2$ and $\eta^2$, we get
\begin{align}
 B^2=\frac{12}{x(9-4\Lambda x)},\quad \eta^2=\frac{3x(4\Lambda x-9)^4}{256(3-\Lambda x)^3(9-6\Lambda x+\Lambda^2x^2)}. \label{eq_NullGeod_Adiscrim}
\end{align}
Then in the space $(B^2,L^2)$, Eq.~\Eqref{eq_NullGeod_Adiscrim} describes a curve parametrised by Eq.~\Eqref{eq_xtor}, where $r=\sqrt{x}$ gives the corresponding radius of the circular orbit.

Setting $\Lambda=0$ to the first equation in \Eqref{eq_NullGeod_Adiscrim} leads to $x=4/3B^2$, or
\begin{align}
 r=\sqrt{x}=\frac{2}{\sqrt{3}B}.
\end{align}
This recovers the circular photon orbit radius of the $\Lambda=0$ Melvin universe \cite{Thorne:1965} in our coordinates.\footnote{The radial coordinate $\rho$ used by Thorne in \cite{Thorne:1965} is related to the present one by $\rho=\half Br$,  so $\rho=\frac{1}{\sqrt{3}}$ is the circular photon orbit radius in Thorne's coordinates.} For $\Lambda\neq0$, solving the first equation in \Eqref{eq_NullGeod_Adiscrim} for $x$ leads to
\begin{align}
 x_\pm=r_\pm^2=\frac{9B\pm\sqrt{81B^2-192\Lambda}}{8B\Lambda}. \label{eq_NullGeod_lightrings}
\end{align}
For $\Lambda<0$ (AdS), only the upper sign yields a real solution for $r$, therefore there is only one branch of circular orbits for the magnetised AdS spacetime. On the other hand, for $\Lambda>0$ (dS) there are two distinct branches of circular orbits for $B^2>\frac{192}{81}\Lambda$.

\subsection*{The $\Lambda>0$ case}

In the dS case, the domain for the radial coordinate is $0<r<\sqrt{3/\Lambda}$, where $r=\sqrt{3/\Lambda}$ is the cosmological horizon. In terms of $x$, the domain is $0<\Lambda x<3$. The expression for $B^2$ in Eq.~\Eqref{eq_NullGeod_Adiscrim} diverges as $\Lambda x\rightarrow\frac{9}{4}$, and beyond that, is negative at $\Lambda x>9/4=2.25$. Therefore there are no circular null geodesics with radii $\frac{9}{4}<x<3$, in the neighbourhood of the horizon.

\begin{figure}
  \centering
  \begin{subfigure}[b]{\textwidth}
    \centering
    \includegraphics[scale=1]{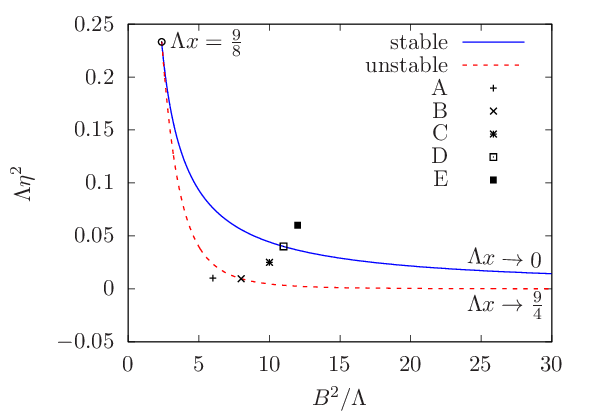}
    \caption{Parameter space.}
    \label{fig_NullGeodesics-dS_ParamSpace}
  \end{subfigure}
  \begin{subfigure}[b]{0.32\textwidth}
    \centering
    \raisebox{0cm}{\includegraphics[width=\textwidth]{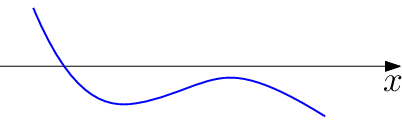}}
    \caption{Point A}
    \label{fig_NullGeodesics-dS_PointA}
  \end{subfigure}
  \begin{subfigure}[b]{0.32\textwidth}
    \centering
    \raisebox{0cm}{\includegraphics[width=\textwidth]{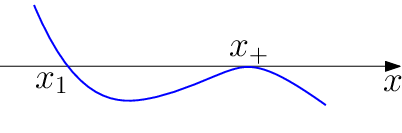}}
    \caption{Point B}
    \label{fig_NullGeodesics-dS_PointB}
  \end{subfigure}
  \begin{subfigure}[b]{0.32\textwidth}
    \centering
    \raisebox{0cm}{\includegraphics[width=\textwidth]{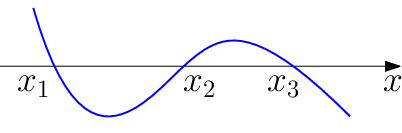}}
    \caption{Point C}
    \label{fig_NullGeodesics-dS_PointC}
  \end{subfigure}
  \begin{subfigure}[b]{0.32\textwidth}
    \centering
    \raisebox{0cm}{\includegraphics[width=\textwidth]{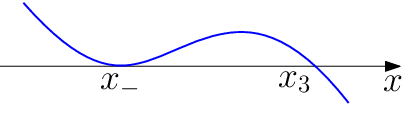}}
    \caption{Point D}
    \label{fig_NullGeodesics-dS_PointD}
  \end{subfigure}
  \begin{subfigure}[b]{0.32\textwidth}
    \centering
    \raisebox{0cm}{\includegraphics[width=\textwidth]{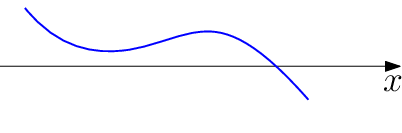}}
    \caption{Point E}
    \label{fig_NullGeodesics-dS_PointE}
  \end{subfigure}
  \caption{Parameter space (a) and sketches [(b)--(f)] of the graphs of $P$ for various values of $(B^2,\eta^2)$ for null geodesics in the dS ($\Lambda>0$) case. The sketches for Points A--E correspond to the representative behaviours around the respective values marked A--E on the parameter space. Points on the solid blue curve give stable circular orbits, while the dashed red curve is the branch of unstable circular orbits. The cusp where the two branches meet are the outermost stable circular orbit (OSCO).}
  \label{fig_NullGeodesics-dS}
\end{figure}

For circular orbits of various other radii $x=r^2$, a plot of Eq.~\Eqref{eq_NullGeod_Adiscrim} in parameter space $(B^2,\eta^2)$ is shown in Fig.~\ref{fig_NullGeodesics-dS_ParamSpace}. The solid blue curve represents circular orbits for $0<\Lambda x<\frac{9}{8}$, coming in from $(B^2,\eta^2)=(\infty,0)$, and ending at the cusp at $(\frac{64}{27}\Lambda,\frac{729}{3125\Lambda})\simeq(2.370\Lambda ,0.233/\Lambda)$, at $\Lambda x=\frac{9}{8}$. Let us call the circular orbit at this radius $r=\sqrt{9/8\Lambda}=3/\sqrt{8\Lambda}$ the \emph{outermost stable circular orbit} (OSCO). Since the circular orbits with radii larger than this are unstable. This is the dashed red curve in Fig.~\ref{fig_NullGeodesics-dS_ParamSpace}. The unstable circular orbits exist for $\frac{9}{8}<\Lambda x<\frac{9}{4}$. As the upper end of this inequality is reached, $(B^2,\eta^2)\rightarrow(\infty,0)$. There are no circular orbits in the neighbourhood of the cosmological horizon $\frac{9}{4}<\Lambda x<3$, where the cosmological horizon is at $x=3=r^2/\Lambda$.

Turning now to other values of $(B^2,\eta^2)$, we start in the region below the branch of unstable circular orbits (below the dashed red curve) in Fig.~\ref{fig_NullGeodesics-dS_ParamSpace}. For $(B^2,\eta^2)$ taking values in this region, the graph of $P$ has one real positive root, as sketched in Fig.~\ref{fig_NullGeodesics-dS_PointA}, and $P$ is negative for $x$ larger than this root, which contains the horizon $\Lambda x=3$. Therefore photons in this case may plunge through the horizon. Varying $(B^2,\eta^2)$ continuously, one reaches the branch of unstable circular orbits, where a degenerate root $x_+$ starts to appear for $P$. The graph in this case is sketched in Fig.~\ref{fig_NullGeodesics-dS_PointB}.

If $(B^2,\eta^2)$ continues into the region between the two curves, say, around Point C in Fig.~\ref{fig_NullGeodesics-dS_ParamSpace}, the graph of $P$ now has three real roots, marked $x_1$, $x_2$, and $x_3$ as sketched in Fig.~\ref{fig_NullGeodesics-dS_PointC}. The domain of motion for null geodesics in this case is $\sqrt{x_1}\leq r\leq\sqrt{x_2}$ and $r\geq\sqrt{x_3}$. Photons in this latter domain might escape beyond the cosmological horizon at $r=\sqrt{3/\Lambda}$, whereas those in the former domain are in bound orbit. This region in particular reflects the opposing effects of $B$ and $\Lambda>0$. It is known that the original Melvin universe acts as a potential well for photons, preventing them from escaping to infinity \cite{Thorne:1965,Karas:1990,Lim:2015oha}. On the other hand, the positive $\Lambda$ in dS-like spacetimes serves as a repulsive gravity. When $B$ is small, but still above $\sqrt{192\Lambda/81}$, the spacetime is approximately Melvin, far away from the cosmological horizon, thus having a potential well $\sqrt{x_1}<r<\sqrt{x_2}$ for photons. Close to the horizon, we have the region $r>\sqrt{x_3}$ in which photons are drawn towards the horizon.

As $(B^2,\eta^2)$ continuously approaches the branch of stable circular orbits from below, the two roots $x_1$ and $x_2$ coalesce into a degenerate root $x_-$, whose expression is according to the lower sign in Eq.~\Eqref{eq_NullGeod_lightrings}. The sketch of $P$ in this case is Fig.~\ref{fig_NullGeodesics-dS_PointD}. Continuing above this point, the degenerate root becomes complex, and once again $P$ is left with one real root, and the only possible orbit is one that may plunge through the horizon.

\subsection*{The $\Lambda<0$ case}

\begin{figure}[htb!]
 \centering
 \begin{subfigure}[b]{\textwidth}
    \centering
    \includegraphics[scale=1]{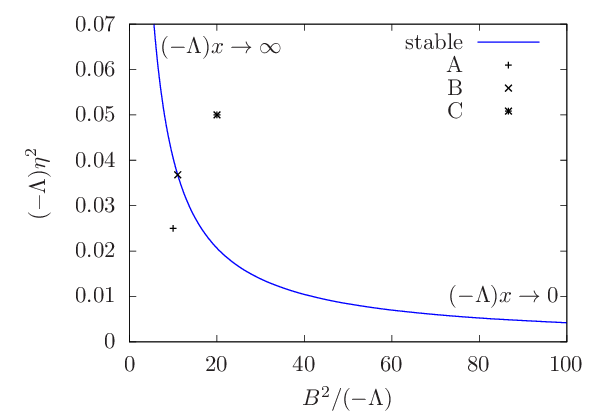}
    \caption{Parameter space.}
    \label{fig_NullGeodesics-AdS_ParamSpace}
  \end{subfigure}
 \begin{subfigure}[b]{0.32\textwidth}
    \centering
    \raisebox{0cm}{\includegraphics[width=\textwidth]{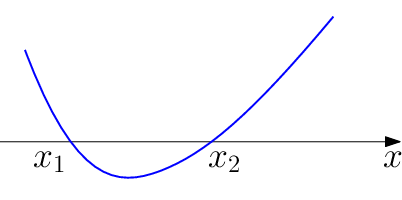}}
    \caption{Point A}
    \label{fig_NullGeodesics-AdS_PointA}
  \end{subfigure}
  \begin{subfigure}[b]{0.32\textwidth}
    \centering
    \raisebox{0cm}{\includegraphics[width=\textwidth]{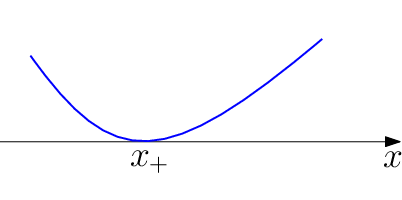}}
    \caption{Point B}
    \label{fig_NullGeodesics-AdS_PointB}
  \end{subfigure}
  \begin{subfigure}[b]{0.32\textwidth}
    \centering
    \raisebox{0cm}{\includegraphics[width=\textwidth]{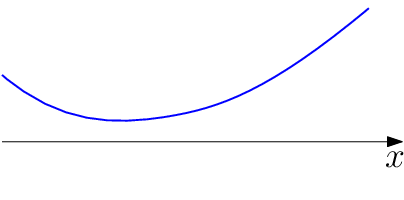}}
    \caption{Point C}
    \label{fig_NullGeodesics-AdS_PointC}
  \end{subfigure}
 \caption{Parameter space for null geodesics in the AdS ($\Lambda<0$) case. The blue curve is the branch of stable circular orbits according to Eq.~\Eqref{eq_NullGeod_Adiscrim}.}
 \label{fig_NullGeodesics-AdS}
\end{figure}

In the AdS case, we have spatial infinity $0<r<\infty$; in terms of $x,$ it is just $0<(-\Lambda)x<\infty.$ The expression $B^2$ in Eq.~\Eqref{eq_NullGeod_Adiscrim} is positive for any values of $x \geq 0.$

For circular orbits of various $x=r^2$, a plot of Eq.~\Eqref{eq_NullGeod_Adiscrim} in parameter space $(B^2,\eta^2)$ is shown in Fig.~\ref{fig_NullGeodesics-AdS_ParamSpace}. As shown, there is only one solid blue curve indicating the set of stable circular orbits coming in from $(B^2, \eta^2)=(\infty, 0).$ It is shown in Fig.~\ref{fig_NullGeodesics-AdS_PointB} that there is exactly one positive root $x_+$ for point B.

We study other values of $(B^2, \eta^2)$. Take point A , for example, which is lying below the curve. For geodesics in this region, $P$ has two positive roots in $x$ where the domain is $\sqrt{x_1} \leq r \leq \sqrt{x_2}$, in bound orbit, as sketched in Fig.~\ref{fig_NullGeodesics-AdS_PointA}. For points above the curve e.g. point C, there are no null geodesics as shown in Fig.~\ref{fig_NullGeodesics-AdS_PointC}.

It is clearly seen that, as we vary parameters $(B^2, \eta^2)$ to cross the curve from below, the two roots $x_1, x_2$ forming the bound orbits collapse into one single $x_+$ as we hit the curve as shown in Fig.~\ref{fig_NullGeodesics-AdS_PointB}. Moving further up away from $P$ results in no geodesics.

\subsection{Time-like geodesics}

As before, since $r$, $B$, $L$, and $E$ all appear in even powers in the effective potential, we let $x=r^2$. The effective potential, where $\epsilon=-1$ for Eq.~\Eqref{eq_EffectiveEnergy}, is
\begin{align}
 U=\frac{Q}{48x(4+B^2x)^2},
\end{align}
where $Q$ is a degree-5 polynomial in $x$,
\begin{align}
 Q&=-\Lambda x^5L^2B^8+B^4(-16\Lambda L^2B^2+3L^2B^4-16\Lambda)x^4+16B^2(3B^2+3L^2B^4-6\Lambda L^2B^2-8\Lambda)x^3\nonumber\\
  &\hspace{1cm}+(288L^2B^4+384B^2-256\Lambda L^2B^2-256\Lambda)x^2\nonumber\\
  &\hspace{1cm}+(768L^2B^2-768E^2-256\Lambda L^2+768)x+768L^2.
\end{align}
The domain of allowed motion, $U\leq 0$ is equivalent to $Q\leq 0$. The condition for circular orbits is
\begin{align}
 Q=Q'=0,
\end{align}
where primes denote derivatives with respect to $x$. Solving these simultaneous equations for $E^2$ and $L^2$, we get
\begin{align}
 L^2&=\frac{16x^2(6B^2-4\Lambda-3\Lambda B^2x)}{(4+B^2x)^2(12-9B^2x+4\Lambda B^2 x^2)},\quad E^2=\frac{(3-\Lambda x)^2(4-B^2x)(4+B^2x)^2}{48(12-9B^2x+4\Lambda B^2 x^2)}, \label{eq_TimelikeGeod_Adiscrim}
\end{align}
which, for a given $x$, describes the required angular momentum and energy for circular orbits of radius $r=\sqrt{x}$. Note that $L^2$ and $E^2$ share a same factor in the denominator, which equals zero when Eq.~\Eqref{eq_NullGeod_Adiscrim} is satisfied, which is the limit $L,E\rightarrow\infty$ to the circular photon orbits.

\subsection*{The $\Lambda>0$ case}

\begin{figure}
 \centering
 \includegraphics{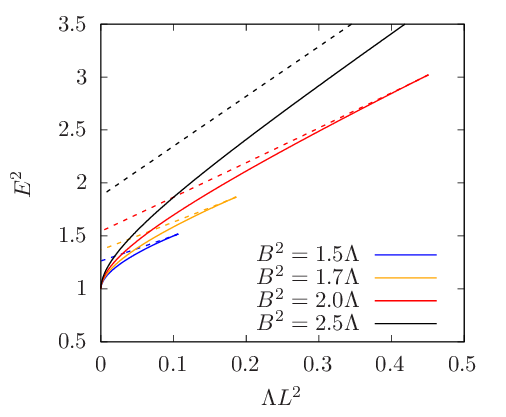}
 \caption{Parameter space for time-like geodesics in the magnetised dS ($\Lambda>0$) case. The solid curves represent the branches of stable circular orbits, whereas the dashed curves represent unstable circular orbits. For $B<\frac{192}{81}\Lambda\simeq2.370\Lambda$, the two branches meet at a cusp, shown here for $B=1.5\Lambda$, $1.7\Lambda$, and $2.0\Lambda$. If $B>\frac{192}{81}\Lambda$, the two branches do not meet. Instead, they extend to infinity, as shown here for $B=2.5\Lambda$.}
 \label{fig_TimelikeGeodesics-dS_ParamSpace}
\end{figure}

In the equation for $L^2$ in \Eqref{eq_TimelikeGeod_Adiscrim}, we know that the right hand side is positive for $0<\Lambda x<3$ provided that $B^2>\frac{2}{3}\Lambda$, so there are no circular orbits for $B^2<\frac{2}{3}$. The plots for a few values of B are shown in Fig.~\ref{fig_TimelikeGeodesics-dS_ParamSpace}. For each $B$, there are two branches, drawn in solid and dashed curves, respectively. The lower branch (solid curves) is the branch of stable circular orbits starting at $(L^2,E^2)=(0,1)$ for $x=0$. Then, two possibilities might happen.

\begin{enumerate}
 \item If $\frac{2}{3}\Lambda<B^2<\frac{192}{81}\Lambda$, the stable branch meets the unstable one at a cusp where $x=x_{\mathrm{OSCO}}$ which we will call the \emph{outermost stable circular orbit} (OSCO). Beyond $x_{\mathrm{OSCO}}$ the curve continues as the unstable branch $x_{\mathrm{OSCO}}<x<\frac{6B^2-4\Lambda}{3\Lambda B^2}$. As can be seen in Fig.~\ref{fig_TimelikeGeodesics-dS_ParamSpace}, the location of the cusp depends on $B$. Increasing $B$ has the cusp moving further from the origin of the $(L^2,E^2)$ plane, until it moves to infinity as $B^2\rightarrow\frac{192}{81}\Lambda$.

 \item For $B^2>\frac{192}{81}\Lambda$, the common denominator in Eq.~\Eqref{eq_TimelikeGeod_Adiscrim} vanishes at the limit
\begin{align}
 x\rightarrow x_\pm=\frac{9B\pm\sqrt{81B^2-192\Lambda}}{8\Lambda B},
\end{align}
where $L^2$ and $E^2$ goes to infinity. These are the limits to the circular null orbits of the previous subsection, where the radii agrees with \Eqref{eq_NullGeod_lightrings}. In any case, the lower branch is the set of stable circular orbits of radii $0<x<x_-$, whereas the upper branch is the unstable ones of radii $x_+<x<\frac{6B^2-4\Lambda}{3\Lambda B^2}$. No circular orbits exist with radii $x_-<x<x_+$, and the stable and unstable branch do not connect.
\end{enumerate}

When $(L^2,E^2)$ take values in the region bounded between the stable and unstable branches, there are three positive roots for $Q$ for which $Q\leq0$ for $x$ between the first two roots, or larger than the third root. In other words, the graph of $Q$ has a similar sketch to Fig.~\ref{fig_NullGeodesics-dS_PointA}, with three real roots $x_1<x_2<x_3$. We then have bound orbtis at $x_1\leq x\leq x_2$, and for $x>x_3$, orbits which plunges through the dS horizon.

On the other hand, when $(L^2,E^2)$ takes values outside the region bounded by the branches, $Q$ has only one real root, and the graphs have sketches similar to Fig.~\ref{fig_NullGeodesics-dS_PointB} or \ref{fig_NullGeodesics-dS_PointC}. In this case, there are no bound orbits, and all time-like particles can plunge through the horizon.

\subsection*{The $\Lambda<0$ case}

Similar to the null geodesics case, we have spatial infinity for AdS space where the domain is $0<(-\Lambda)x<\infty.$ Looking at the common denominator of $L^2$ and $E^2$ in Eq.~\Eqref{eq_TimelikeGeod_Adiscrim}, both $L^2$ and $E^2$ diverge at
\begin{align}
 x_+ = \frac{9B-\sqrt{81B^2-192\Lambda}}{8\Lambda B},
\end{align}
suggesting that the domain for circular orbits is $0<x<x_+$. When $x \rightarrow x_+, L^2 \rightarrow \infty$ which is the limit to \Eqref{eq_NullGeod_lightrings} of the circular null geodesics. 
Any trajectory beyond $x_+$ yields $L^2<0$, which is impossible. When $B \rightarrow \infty, \, x_+ \rightarrow 0.$ That means the stronger the magnetic field gets, the heavier it compresses the region where stable circular orbits can exist.

\begin{figure}[htb!]
 \centering
 \includegraphics{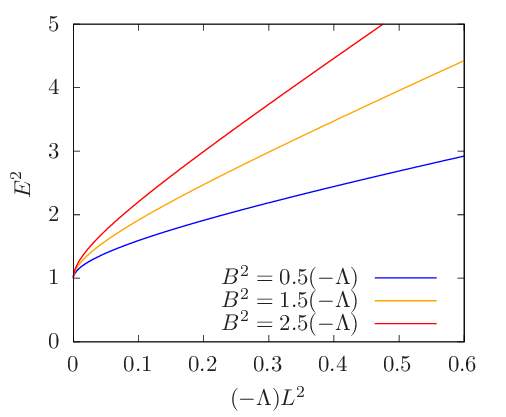}
 \caption{Parameter space for time-like geodesics in the AdS ($\Lambda<0$) case for various values of $B$. The curves for each $B$ are the branches of circular orbits, all of which are stable. Values of $(L^2,E^2)$ lying above the curve give bounded orbits, and those below the curve has no time-like geodesics, as $Q>0$ for all $x>0$.}
 \label{fig_TimelikeGeodesics-AdS_ParamSpace}
\end{figure}

As illustrated in Fig.~\ref{fig_TimelikeGeodesics-AdS_ParamSpace}, there is only one branch of stable circular orbits for each $B.$ The branches start at $(L^2, E^2)=(0,1)$ for $x=0.$ When $(L^2, E^2)$ takes values in the region above the stable branches, there are two positive roots for $Q$ for which $Q \leq 0.$, similar to sketch Fig.~\ref{fig_NullGeodesics-AdS_PointA}. In other words, there exists a bound orbit $x_1 \leq x \leq x_2$ for each $B$. There are no time-like geodesics for particles in regions below the stable branches, as $Q \geq 0$ for all $x \geq 0.$

\section{Conclusion} \label{sec_conclusion}

In this work, we have obtained dS- and AdS-like counterparts to the Melvin magnetic universe, in the sense that the seed spacetime is the (A)dS with the parameter $\Lambda$, which is called as such because we have been working within the Einstein--Maxwell--fluid model where, instead of a cosmological constant term, we had an anisotropic fluid whose energy and pressure reduces to an effective cosmological constant in the zero field limit. In other words, when the magnetic field $B$ is set to zero, we recover the (A)dS spacetimes. For $B\neq0$, the energy density and pressures are given in Eq.~\Eqref{4d_rho_p}. When $\Lambda$ is set to zero, we recover the Melvin universe.

We first approached this problem in the weak field regime by solving Maxwell's equation in a fixed (A)dS background. In the vacuum case we obtained magnetic multipole analogues to the electric ones of \cite{Herdeiro:2015vaa}. The Killing vector method of Wald \cite{Wald:1974np} could also be applied here, but as the background is not Ricci-flat, a current source in the Maxwell equation is required. The gauge potential from the latter method is extended to a strong field case, where a modified Harrison transformation is applied in Einstein--Maxwell--fluid gravity, where an anisotropic fluid is present.

In Sec.~\ref{sec_geod}, we only considered equatorial geodesics. But still the geodesics already show a rich behaviour due to the interplay between the physics of (A)dS and the magnetic field. In particular, the Melvin-type magnetic field serves as a potential well \cite{Thorne:1965,Karas:1990,Lim:2015oha}, and so does AdS spacetime (when $\Lambda<0$). On the other hand, the dS spacetime ($\Lambda>0$) gives a repulsive force that pulls particles away from the origin, towards the horizon. This can be seen with the null geodesics, where in the dS case results in two circular orbits, one stable and another unstable. The stable circular orbits have smaller radii than the unstable ones, which can be understood since regions at small radii are far from the horizon and its dS physics, so the dominant effect would be the potential well of the magnetic field. These rich behaviours were obtained just from studying the properties of the (numerators of) the effective potentials. But since the effective potential is a rational function, it might be possible that analytical solutions to the geodesic equations may be obtained using the methods of \cite{Hackmann:2008tu,Hackmann:2008zz}.


As a final remark, in this paper we have focussed exclusively on Melvin-type solutions. That is, magnetic universes in the absence of black holes. A Schwarzschild-type black hole can easily be added to the solution \Eqref{4d_soln}, by adding a term to $f$, as
\begin{align}
 f=1-\frac{2M}{r}-\frac{\Lambda}{3}r^2.
\end{align}
With $M>0$, we will have a Schwarzschild--(Anti-)de Sitter-type solution \cite{Kottler:1918cxc}, with a black hole horizon, along with the curvature singularity behind it. With this $f$ above, the solution \Eqref{4d_soln} still solves the field equations, and may perhaps be regarded as the (A)dS generalisation of the Ernst solution \cite{Ernst:1976mzr}. In fact, since the Harrison-like transformation can be applied to non-vacuum solutions with an anisotropic fluid, other more general black holes can be magnetised, such as in \cite{Sajadi:2025prp}. 

\section*{Acknowledgments}
Y.-K.~L is supported by Xiamen University Malaysia Research Fund (Grant no. XMUMRF/ 2021-C18/IPHY/0011).

\newpage
\bibliographystyle{mdS}

\bibliography{mdS}

\end{document}